\documentclass[a4paper,twocolumn,11pt]{quantumarticle}
\pdfoutput=1

\usepackage{bbold}
\usepackage{amsmath}
\usepackage{mathtools}
\usepackage{graphicx}

\usepackage{multirow}
\usepackage[table,xcdraw]{xcolor}
\usepackage[colorlinks=true, allcolors=blue]{hyperref}

\newcommand{\edges}{{\mathcal{E}}}
\newcommand{\vertices}{{\mathcal{V}}}

\newcommand{\ket}[1]{\left|#1\right\rangle}

\newcommand{\argmax}{\text{argmax}}
\newcommand{\argmin}{\text{argmin}}
\newcommand\xor{\oplus}
\usepackage{amsmath,amssymb}
\usepackage[a4paper, portrait,text={19cm,26cm},centering]{geometry}
\usepackage[most]{tcolorbox}
\usepackage{mathrsfs}
 \usepackage{graphicx}

\usepackage[utf8]{inputenc}
\usepackage{graphicx,stackrel}
\usepackage{dcolumn}
\usepackage{bm}
\usepackage[bbgreekl]{mathbbol}
\usepackage{amsmath,dsfont}
\usepackage[normalem]{ulem}
\usepackage{color}
\usepackage[ruled,vlined]{algorithm2e}
\usepackage[utf8]{inputenc}
\usepackage{subcaption}

\usepackage{hyperref}
\usepackage{multirow}
\usepackage{caption}
\usepackage{ragged2e}

\begin{document}

\title{Efficient protocol for solving combinatorial graph problems on neutral-atom quantum processors}
\author{Wesley da Silva Coelho}
\affiliation{PASQAL SAS, 7 rue Léonard de Vinci, 91300 Massy, France}

\author{Mauro D'Arcangelo}
\affiliation{PASQAL SAS, 7 rue Léonard de Vinci, 91300 Massy, France}

\author{Louis-Paul Henry }
\email{louis-paul.henry@pasqal.com}
\affiliation{PASQAL SAS, 7 rue Léonard de Vinci, 91300 Massy, France}


\begin{abstract}
On neutral atom platforms, preparing specific quantum states is usually achieved by pulse shaping, {\sl i.e.}, by optimizing the time-dependence of the Hamiltonian related to the system. This process can be extremely costly, as it requires sampling the final state in the quantum processor many times. Hence, determining a {\sl good} pulse, as well as a {\sl good} embedding, to solve specific combinatorial graph problems is one of the most important bottlenecks of the analog approach. In this work, we propose a novel protocol for solving hard combinatorial graph problems that combines variational analog quantum computing and machine learning. Our numerical simulations show that the proposed protocol can reduce dramatically the number of iterations to be run on the quantum device. Finally, we assess the quality of the proposed approach by estimating the related Q-score, a recently proposed metric aimed at benchmarking QPUs.
\end{abstract}

\maketitle


\section*{Introduction} 

A lot of effort is currently being put into designing quantum algorithms and hardware that could provide an advantage over classical computers. This advantage can take the form of more accurate results, a faster convergence, or even a lower energy consumption.
These solutions are developed on very different platforms, using a wide range of technologies. The most prominent ones are based on trapped ions~\cite{PRXQuantum.2.020343, Benhelm_2008}, Josephson junctions~\cite{Kjaergaard2020, Devoret2004} and Rydberg neutral atoms~\cite{henriet2020quantum, wurtz2022}. In each case, the information is stored in a two-level system constituting the qubits. Different sets of quantum gates~\cite{Qgates} can then be implemented and, for a given quantum algorithm, the effective quantum circuits can vary significantly across platforms. 
Additionally, there are problems for which even the Noisy Intermediate Scale Quantum (NISQ) processors~\cite{Preskill_2018} are expected to provide an advantage. This could be obtained from an {\it analog} approach where, as opposed to the case of {\it digital} quantum computing, the quantum operations are not divided into discrete consecutive steps (gates), but are rather the result of a time-dependent control of the Hamiltonian acting upon the qubits. 
This solution will be very intrinsically problem- and platform-specific, further complicating any comparisons.

Hence, comparing different approaches can be difficult, and it is hard to define a metric that can be applied to all of them, including classical ones. People have used many different ones, from the bare number of qubits available to the more involved {\sl Quantum Volume}~\cite{PhysRevA.100.032328}, but none seems to be universally fair. To overcome the aforementioned problems, a new metric called \textit{Q-score} was recently introduced by Atos~\cite{qscore}, fitting this very need. It consists in quantifying the performances of a given device or method in solving a specific combinatorial optimization problem, such as the Maximum Cut (MaxCut). 

Graphs are used in a vast spectrum of fields. In particular, several combinatorial problems either are or can be defined on graphs.  
Those problems are of particular relevance for Quantum Computing (QC). It is particularly the case for NISQ-era platforms~\cite{Bharti_2022, Nishi_2021}. Indeed, they are very well suited for measurement-based computing, that may be combined with adiabatic annealing. This type of QC is particularly robust to noise (noise can even be an advantage~\cite{Novo_2018}), as there is a direct correspondence between the state of the computational basis in which the qubits are measured and the solution to the graph problem.

Among the quantum computing platforms, neutral atoms are particularly well suited to solving these combinatorial graph problems~\cite{graph_rydberg1, graph_rydberg2, graph_rydberg3, dalyac2021qualifying}. In fact, the Ising Hamiltonian describing the dynamics of the qubits is closely related to the cost function to be minimized. 
Solving the problems is then equivalent to finding the ground state of the system, which can be achieved by adiabatic annealing~\cite{QA1}, as it has been shown in the case of the Maximum Independent Set (MIS) problem~\cite{dalyac2021qualifying, graph_rydberg2}. It is worth mentioning that, in this case, one does not necessarily need to prepare the exact ground state, but only needs to prepare a state with a sufficient overlap with it. In particular, this allows extending the method to cases where the Hamiltonian only partially reproduces the cost function. State preparation is usually achieved by pulse shaping, {\sl i.e.} by optimizing the time-dependence of the Hamiltonian. This process can be extremely costly, as it requires a lot of sampling of the final state in the quantum processor (or in its emulator).

\subsection*{Our contributions}
 In this paper, we propose a Machine Learning protocol to predict the shape of the Hamiltonian and therefore reduce dramatically the number of iterations to be run on the Quantum device. Once trained on a training set consisting of graphs and their associated solving pulses, the Machine Learning model is then able to provide a good pulse for any new, unseen instance of the problem without any further training or optimization process. We also present different strategies to create neutral-atom registers that are specifically tailored to different graph classes.

This paper is structured as follows : 
we introduce the different combinatorial graphs problems in Section \ref{background}, then describe how they are particularly relevant for neutral atom platforms in Section \ref{rydberg}. The methods are detailed in Section \ref{methods} and the results analyzed in Section \ref{results}.

\section{Background} \label{background}
Graph Theory~\cite{bondy1976graph} has been widely studied by both industrial and academic communities, and has a vast range of applications on several real-world systems. For instance, graphs can be used to encode telecommunication networks~\cite{9625601}, social experiments~\cite{roth2010suggesting}, and physical systems~\cite{barahona1988application}.  Graphs are data structures composed of a set of elements called \textit{vertices} (also known as nodes) that can potentially be connected. These connections are called \textit{edges} and can encode different information, such as the distance between their endpoints or the importance of such connections. Formally, a graph $G = (\vertices,\edges)$ is composed of a set of vertices $\vertices$ and edges $\edges$ seen as unordered pairs of vertices $\{i,j\} \in \vertices^2$ representing the existence of a connection between  $i$ and $j$. In the following, we present two important combinatorial graph problems.

\subsection{Combinatorial graph problems}
The Maximum Cut (MaxCut) and Maximum Independent Set (MIS) problems belong to the well-known class of NP-Complete problems~\cite{karp1972reducibility}, and their optimization versions have been studied in depth. While the MaxCut problem is equivalent to minimizing the Hamiltonian of a spin glass~\cite{barahona1988application}, the solutions of the MIS problem on unit-disk graphs can be encoded as the ground state of the Hamiltonian describing neutral-atoms devices~\cite{henriet2020quantum}. In the following, we formally define these problems and present their related Quadratic Unconstrained Binary Optimization (QUBO) formulations.

\subsubsection{Problem definitions}

\textit{Maximum independent set problem}: Given a  graph $G = (\vertices, \edges)$, an independent set is a subset of vertices $\tilde\vertices \subset \vertices$ such that no two elements of $\tilde\vertices$ are connected by an edge. The independent sets of $G$ can formally be defined as follows:
    \begin{equation}
        IS_G = \left\{\tilde\vertices\subset\vertices \bigm| \ \tilde\vertices^2\cap\edges = \emptyset\right\}
    \end{equation}
where $\tilde \vertices^2$ are all the possible edges connecting the vertices in $\tilde \vertices$. The Maximum Independent Set is therefore the largest element of $IS_G$:
    \begin{equation}
        \text{MIS}(G) = \underset{\tilde\vertices\in IS_G}{\argmax} \;|\tilde\vertices|.
    \end{equation}
\textit{Maximum cut problem}: Given a  graph $G = (\vertices, \edges)$, a cutting is a partition of $\vertices$ into two disjoints sets $\tilde\vertices\subset\vertices$ and $\vertices'= \vertices\symbol{92}\tilde\vertices$. Associate to the cutting set $\tilde\vertices$ the subset of edges $\tilde\edges \subset \edges$ connecting $\tilde\vertices$ to $\vertices'$:
\begin{equation}
    \tilde\edges = \left\{\{i,j\} \in \edges \bigm| \ i \in \tilde\vertices  \xor  j \in \tilde\vertices\right\}
\end{equation}
where $\xor$ represents the logic operator \textit{exclusive or}. Denoting the cut-set $CS_G$ to be the set of all possible $\tilde \edges$, the Maximum Cut (MaxCut) can be defined as:  
\begin{equation}
   \text{MaxCut}(G) =  \underset{\tilde\edges\in CS_G}{\argmax} \;|\tilde\edges|.
\end{equation}

\begin{figure}[t]
\centering
\captionsetup{justification=Justified}
\includegraphics[width=0.20\textwidth, clip=false]{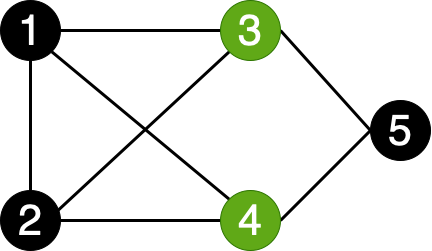}
\caption{Example of solutions for both MaxCut and MIS problem: the set of green nodes represents optimal solutions for both problems.}
\label{mismcex}
\end{figure}

Fig.~\ref{mismcex} depicts an example of a solution for both the MaxCut and MIS problems, given by choosing $
\tilde \vertices$ to be the green nodes. Notice that an MIS or MaxCut solution is not always unique. Black nodes in Fig.~\ref{mismcex} are also an optimal solution for the MaxCut as they cut 6 edges. Similarly, $\{1,5\}$ and $\{2,5\}$ are other optimal solutions for the related MIS problem. 

\subsubsection{Related work}
Solutions to combinatorial problems such as the MIS and MaxCut are traditionally found using classical methods based on approximation algorithms~\cite{berman1994approximating,goemans1994879}, exact approaches~\cite{xiao2017exact,rendl2010solving}, and heuristics techniques~\cite{das2012heuristics,kim2019comparison}. However, recent years were marked by growing interest in emerging quantum computing platforms aimed at solving combinatorial problems, an ideal playground for testing and validating noisy near-term devices. A Quantum Approximate Optimization Algorithm (QAOA) to address combinatorial problems was first proposed by Farhi {\sl et al}~\cite{farhi2014quantum} in 2014. The authors studied the MaxCut problem and showed that the quality of final solutions improves as the unitary gate-based circuit's depth \textit{p} increases. Since then, several works showed the performance of applying QAOA approaches to address combinatorial graph problems on different quantum computers~\cite{brandao2018fixed, zhou2020quantum, crooks2018performance, medvidovic2021classical, herrman2021impact}. For instance, by simulating several gates in parallel, authors in~\cite{crooks2018performance} showed that one can decrease the runtime for solving the MaxCut problem despite limitations on qubit connectivity (at least for small graphs). Moreover, Herrman {\sl et al}~\cite{herrman2021impact} discussed the impact of graph structures for QAOA on MaxCut and presented some predictors of QAOA success related to graph density, odd cycles, and symmetries. More practically, Dalyac {\sl et al}~\cite{dalyac2021qualifying} addressed a real-world problem related to the scheduling of load time intervals within groups of electrical vehicles. By reducing the MIS to the aforementioned problem, the authors showed that the proposed Rydberg atom array-based QAOA can exceed approximation rates of 0.95 after seven layers of the algorithm.

Although most QAOA applications focus on gate-based models of quantum computing, a promising avenue for noisy devices is represented by analog variational algorithms. The analog mode of operation involves the evolution of a quantum system under a continuously controllable resource Hamiltonian rather than the discrete application of a fixed set of quantum gates. Whereas the successful implementation of a gate-based algorithm is limited by the absence of error correction on current devices, an analog algorithm is intrinsically more resilient to noise~\cite{Markovi__2020,ZAK19991583}. In this framework, the role of Rydberg atom arrays~\cite{henriet2020quantum} is recognized as a prominent example of how the ground state of a quantum Hamiltonian directly maps to the solution of a hard combinatorial graph problem, MIS on unit-disk graphs in this case. For instance, as a follow-up to an earlier study~\cite{Pichler}, authors in~\cite{ebadi2022quantum} investigate different analog quantum algorithms driven by closed-loop parameter optimization in order to solve MIS on unit-disk graphs with hundreds of nodes, showing that the number of local minima and the solution degeneracy control the hardness of the problem. 

The quantitative argument linking the Hamiltonian of analog quantum devices and combinatorial graph problems goes through the QUBO formulation of the latter, which is presented next. 

\subsubsection{QUBO formulations}
The MaxCut and MIS problems can be alternatively described in terms of their QUBO (quadratic unconstrained binary optimization) formulations. 
Consider a graph $G=(\vertices,\edges)$ with vertex set $\vertices = \{1, \ldots, M\}$ and edge set $\edges \subset \vertices \times \vertices$. For each vertex $i$ of the graph, let $x_i$ be a binary variable that holds 1 if it is activated in the final solution, and 0 otherwise. The binary vector $\mathbf{x} = \{ x_1, \ldots, x_M\}$ can then be put in one-to-one correspondence with partitions $\tilde \vertices$ of the vertex set $\vertices$ via the identification:
\begin{equation}
    \tilde \vertices (\mathbf{x}) = \left\{i \in \vertices \bigm| \ x_i = 1 \right\}
\end{equation}
The solution to the maximum independent set problem is then given by:
\begin{equation}\label{eq:misqubo}
\hspace{-0.3cm}    \text{MIS}(G) = \underset{\mathbf{x} \in \{0,1\}^M}{\argmin} \left( -\sum_{i \in \vertices} x_i + \sum_{\{i,j\} \in \edges} x_i x_j \right)
\end{equation}
while for the maximum cut problem:
\begin{equation}
   \hspace{-0.275cm} \text{MaxCut}(G) = \underset{\mathbf{x} \in \{0,1\}^M}{\argmin} \left( -\sum_{\{i,j\} \in \edges} (x_i - x_j)^2  \right)
\end{equation}
A little algebra on the QUBO formulation of MaxCut shows that, denoting $N(i)$ the number of neighbors of vertex $i$ (i.e. the numbers of vertices connected to $i$ by an edge), the maximum cut \textit{MaxCut(G)} is equivalently given by: 
\begin{equation}\label{eq:MaxCutqubo}
   \hspace{-0.3cm}  \underset{\mathbf{x} \in \{0,1\}^M}{\argmin} \left( -\sum_{i \in \vertices} N(i) \ x_i + \sum_{\{i,j\} \in \edges} x_i x_j \right)
\end{equation}
The QUBO cost functions of MIS and MaxCut are directly related to the energy of classical Ising (spin glass) models, which have been extensively studied in the mathematics and physics literature~\cite{Barahona1982,yaacoby2022comparison,coja2022ising,zhang2020computational}. Ising models are often written in terms of variables $s_i$ that take value in $\{+1,-1\}$, but the two formulations are related to each other by a linear change of variables $x_i = (s_i +1)/2$ that preserves the order of the interactions. The solution to a QUBO is then translated to minimizing the energy of a physical system, in the spirit of physics-inspired approaches to optimization problems~\cite{Fu1986, Mezard1986, Hartmann2005}. The quantum version of a classical Ising model uses the same Hamiltonian, but the binary variables $s_i$ and $x_i$ are replaced respectively by the Pauli $\hat{\sigma}_z$ operator and the number operator $\hat{n}_i = (\hat{\sigma}_z + \mathds{1})/2$. For example, taking the MIS cost function (\ref{eq:misqubo}) and replacing $x_i \rightarrow \hat{n}_i$ gives:

\begin{equation}
     -\sum_{i \in V} \hat{n}_i + \sum_{\{i,j\} \in E} \hat{n}_i \hat{n}_j.
\end{equation}

Several controllable quantum devices are available today that naturally implement a quantum Ising Hamiltonian. For instance, the Hamiltonian of a system of $M$ Rydberg atoms coupled to a global driving laser with Rabi frequency $\Omega(t)$ and detuning $\Delta(t)$ at instant $t$ reads:
\begin{equation}\label{eq:hamiltonian}
    H(t) = \Omega(t)\sum_{i=1}^M\hat{\sigma}^x_i - \Delta(t)\sum_{i=1}^M \hat{n}_i +\sum_{i<j=1}^{M} U_{ij} \hat{n}_i \hat{n}_j
\end{equation}
where the interaction strength $U_{ij}$ is a function of the distance between atom $i$ and atom $j$. 
\section{Neutral atom QPUs} \label{rydberg}
                  
The spin Hamiltonian \eqref{eq:hamiltonian} can be implemented in arrays of neutral atoms where the $\ket{0}$ and $\ket{1}$ states are atomic energy levels. Typically, $\ket{0}$ is chosen to be the ground state, while $\ket{1}$ represents a highly excited Rydberg state. Rydberg states have strong interactions that decay as a function of the distance $r$ between the atoms (as $r^{-6}$ in the case of $S$ Rydberg states), while the transverse and longitudinal fields $\Omega$ and $\Delta$ correspond to amplitude and detuning of a driving laser that addresses the $\ket{0} \rightarrow \ket{1}$ transition. The atoms are then cooled and trapped in a system of optical tweezers, which allows to arrange them in arbitrary 2D configurations.

Having complete control over the position of every single atom and on the laser parameters opens up the possibility of probing non-trivial quantum dynamics in systems of spins arranged in graph-like structures. The distribution of certain observables for neutral atom systems subject to time-dependent Hamiltonians was used for instance in \cite{Henry_2021} as a fingerprint of the graph itself in the construction of a graph kernel. The so-called Quantum Evolution Kernel was shown to be on par with (if not outperforming) state-of-the-art graph classifiers on real datasets.

The focus of the present study is to represent graphs as systems of Rydberg atoms and to find time-series of control parameters $\Omega(t)$ and $\Delta(t)$ for the Hamiltonian \eqref{eq:hamiltonian} such that the outcome of quantum evolution and measurement is, with high probability, a good solution to the related QUBO representing an instance of the MaxCut or MIS problem.

\subsection{Graph embedding}
\begin{figure}[t] 
    \centering
  \begin{subfigure}[b]{0.5\linewidth}
    \centering
    \includegraphics[width=.8\linewidth]{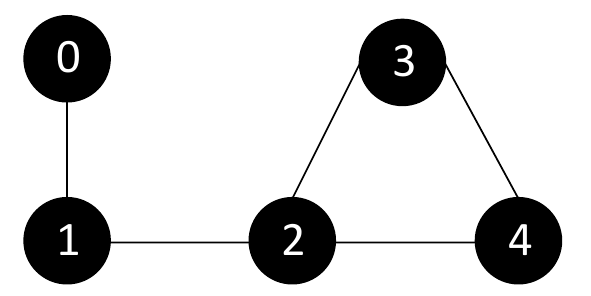} 
    \caption{\centering UD embedding.} 
    \label{fig:udgraphs_a} 
  \end{subfigure}
  \begin{subfigure}[b]{0.5\linewidth}
    \centering
    \includegraphics[width=.65\linewidth]{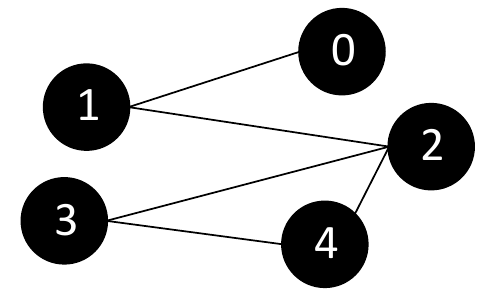} 
    \caption{Non-UD embedding.} 
    \label{fig:udgraphs_b} 
  \end{subfigure} \captionsetup{justification=Justified}
  \caption{(a) and (b) show two different embeddings for the same graph in 2D Euclidean space. The embedding on the left shows that the graph is a unit disk graph.}
  \label{fig:udgraphs} 
\end{figure}

One of the features of neutral atom devices that makes it interesting for combinatorial graph problems is that the ground state of the Hamiltonian can encode exactly the solution to MIS on unit disk (UD) graphs. Given a graph, if an embedding in two dimensions can be found such that two nodes are connected if and only if their Euclidean distance is less than a certain threshold, then the graph is said to be a unit disk graph. An example of a UD graph is given in Fig.~\ref{fig:udgraphs}, where two different embeddings for the same graph are shown: one that makes explicit the unit disk nature of the graph (Fig.~\ref{fig:udgraphs_a}), and one that does not (Fig.~\ref{fig:udgraphs_b}). 

When a UD embedding of a graph is replicated on a neutral atom device, the two-body interaction term in (\ref{eq:hamiltonian}) forbids the simultaneous excitation of two atoms that are closer than a certain distance, a phenomenon known as the Rydberg blockade~\cite{henriet2020quantum}. This ensures that the evolution of the quantum system is restricted to a subspace of the complete Hilbert space where the excitations correspond to independent sets of the graph. For positive detunings, moreover, excitations are energetically favored, leading to the ground state of the system to be a maximum independent set of the graph.

A standard proposal for finding the ground state of a time-dependent Hamiltonian $H(t)$ relies on the adiabatic theorem~\cite{Farhi2001}. With a parametrization of time such that $t\in[0,1]$, assume that $H(0)$ corresponds to a simple Hamiltonian whose ground state can be prepared easily, and $H(1)$ is the Hamiltonian whose ground state one wants to find. The adiabatic theorem ensures that if the system is prepared in the ground state of $H(0)$ and the parameter $t$ is changed slowly enough, then the system will persist in the instantaneous ground state of $H(t)$ for all $t$, and therefore it will eventually find itself in the ground state of $H(1)$. In practice, however, the adiabatic theorem is hard to apply. The energy gap between the ground state and the first excited state typically becomes exponentially small during the quantum evolution~\cite{exponentialgap}. In those situations, for the adiabatic theorem to hold true the quantum evolution is required to be too slow to yield any real advantage.

\subsection{Pulse shaping}
\begin{figure}[b]
\centering
\includegraphics[width=0.45\textwidth]{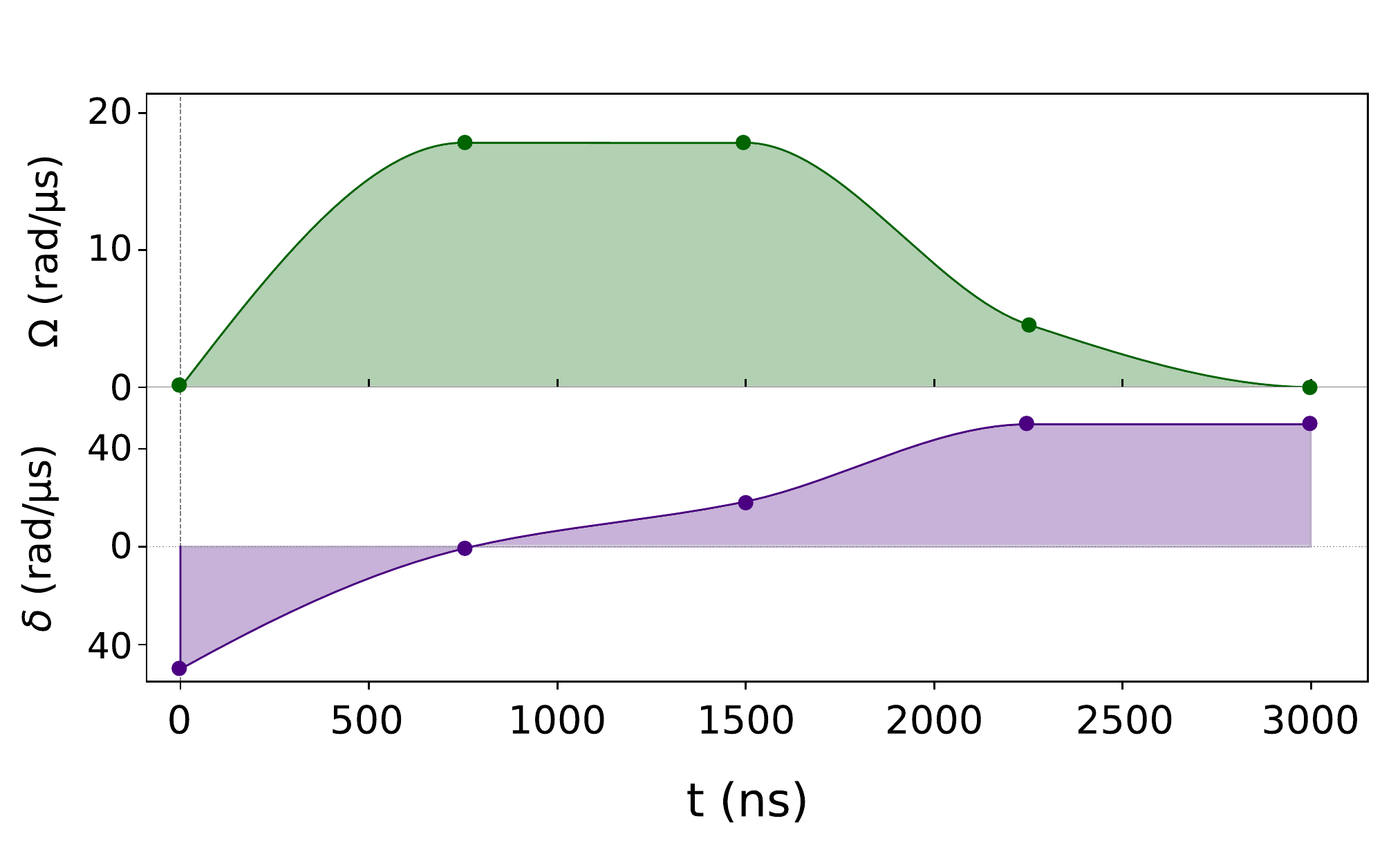}\captionsetup{justification=Justified}
\caption{Best pulse found by the pulse shaping routine for the UD graph depicted in Fig.~\ref{fig:easy_graph}. The top green curve is the value of the Rabi frequency, while the bottom curve is the value of the detuning. As expected, the pulse resembles a simple adiabatic protocol.}
\label{fig:easy_pulse}
\end{figure}
An alternative proposal based on a hybrid quantum-classical approach consists in finding an (in general non-adiabatic) optimal path $$\gamma: t \mapsto (\Omega(t), \Delta(t))$$ in the two-dimensional parameter space of the Hamiltonian. Typically, $\Omega(t)$ and $\Delta(t)$ are parametrized by a few equally spaced points between which the path is smoothly interpolated as shown in Fig.~\ref{fig:easy_pulse}. Since this procedure is aimed at finding the best shape for the laser pulses that drive the quantum system, it will be hereafter referred to as \textit{pulse shaping}. The optimal values for the interpolating points are found using a Bayesian search routine aimed at minimizing a certain objective function. There are a few inequivalent ways of building the objective function, with some strategies working better than others depending on the problem, but in general, they all rely on a sampling of the quantum state that results from the evolution of the system. The evolution of a quantum system is governed by its Hamiltonian $H(t)$, and in general one can say that there exists a mapping parametrized by $\Omega(t)$ and $\Delta(t)$ that brings a system prepared in an initial state $\ket{\psi_0}$ to a certain final state $\ket{\psi_t}$:
\begin{equation}
    \ket{\psi_0} \xmapsto{\Omega(t), \Delta(t)} \ket{\psi_t}.
\end{equation}
If the system is comprised of $M$ atoms, the final state will be in general a normalized superposition of basis states that are in one-to-one correspondence with binary bitstrings of length $M$:
\begin{equation}
    \ket{\psi_t} = \sum_{i=1}^{2^M} a_i \ket{b_i}
\end{equation}
with $\sum_i |a_i|^2 = 1$ and
\begin{equation}
   \hspace{-0.2cm} \ket{b_i} = \ket{b_i^1} \otimes \ldots \otimes \ket{b_i^M},  \quad \ket{b_i^j} = \ket{0} \ \text{or} \ \ket{1}.
\end{equation}

Perfect knowledge of the quantum state $\ket{\psi_t}$ (hence perfect knowledge of the coefficients $a_i$) would require an exponential amount of resources as the system scales up in size. The state is therefore only known approximately through repeated measurements. A single measurement of the state $\ket{\psi_t}$ can be seen as extracting one of the bistrings $b_i$ with probability $|a_i|^2$. Collecting $N$ samples of the state results then in a collection of pairs $$\{(b_i, w_i^{(N)})\}_{i=1,\ldots,2^M}$$ where $w_i^{(N)}$ indicates how many times the bitstring $b_i$ was measured out of $N$ tries. Clearly, one has:
\begin{equation}
    \lim_{N \to \infty} \frac{w_i^{(N)}}{N} = |a_i|^2.
\end{equation}
\begin{figure}[t] 
    \centering \hspace{-0.8cm}
  \begin{subfigure}[b]{0.35\linewidth}
    \centering
    \includegraphics[width=\textwidth]{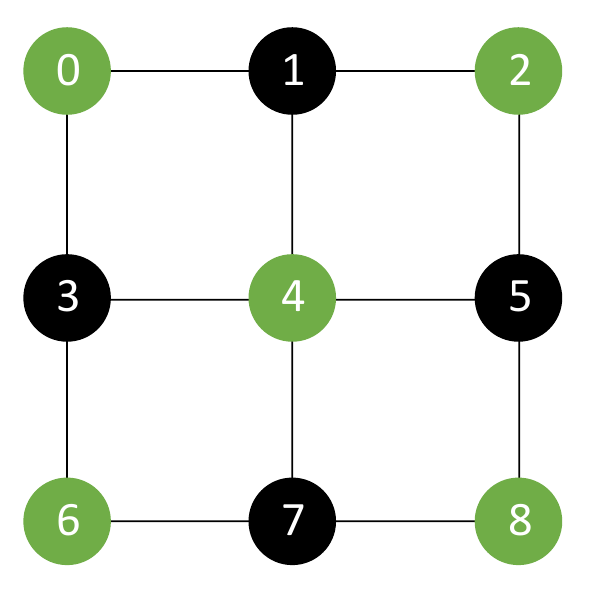} 
    \caption{Unit disk graph.}
    \label{fig:easy_graph} 
  \end{subfigure}\hspace{0.3cm}
  \begin{subfigure}[b]{0.35\linewidth}
    \centering
    \includegraphics[width=\textwidth]{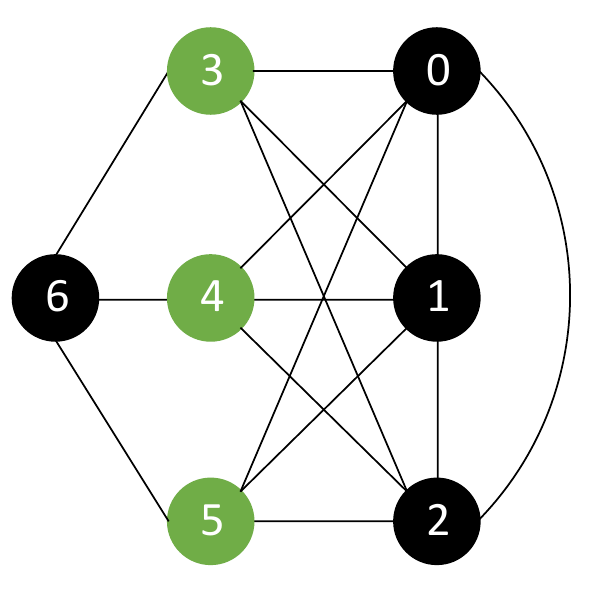}
    \caption[width=\textwidth]{Non-UD graph.}
    \label{fig:hard_graph} 
  \end{subfigure} \captionsetup{justification=Justified}
    \caption{Examples of (a) an {\sl easy} and (b) a {\sl hard} graph for a neutral atom platform. The MIS is indicated in green.}
    \label{fig:graphs} 
\end{figure}
Each bitstring $b_i$ corresponds to a unique bipartition of nodes in the graph that the Rydberg atoms represent, and therefore to each bitstring $b_i$ one can associate a cost given by the QUBO formulation of a combinatorial problem (e.g., MIS \eqref{eq:misqubo} and MaxCut \eqref{eq:MaxCutqubo}). By denoting this cost $C(b_i)$,  three possible ways of defining the objective function to minimize in the Bayesian optimization routine are:
\begin{align}
    \sum_i w_i\ &C(b_i) \label{eq:avgcost}\\
    \min_i\ &C(b_i) \\
    \max_i\ &C(b_i).
\end{align}
The three choices are aimed at finding pulses that produce a final state where respectively:
\begin{enumerate}
    \item good solutions to the QUBO are sampled more frequently;
    \item at least one of the sampled bitstrings is a good solution to the QUBO;
    \item the sampled bitstring with the worst score is still a good solution to the QUBO.
\end{enumerate}

\begin{figure}[t]
\centering
\includegraphics[width=0.5\textwidth]{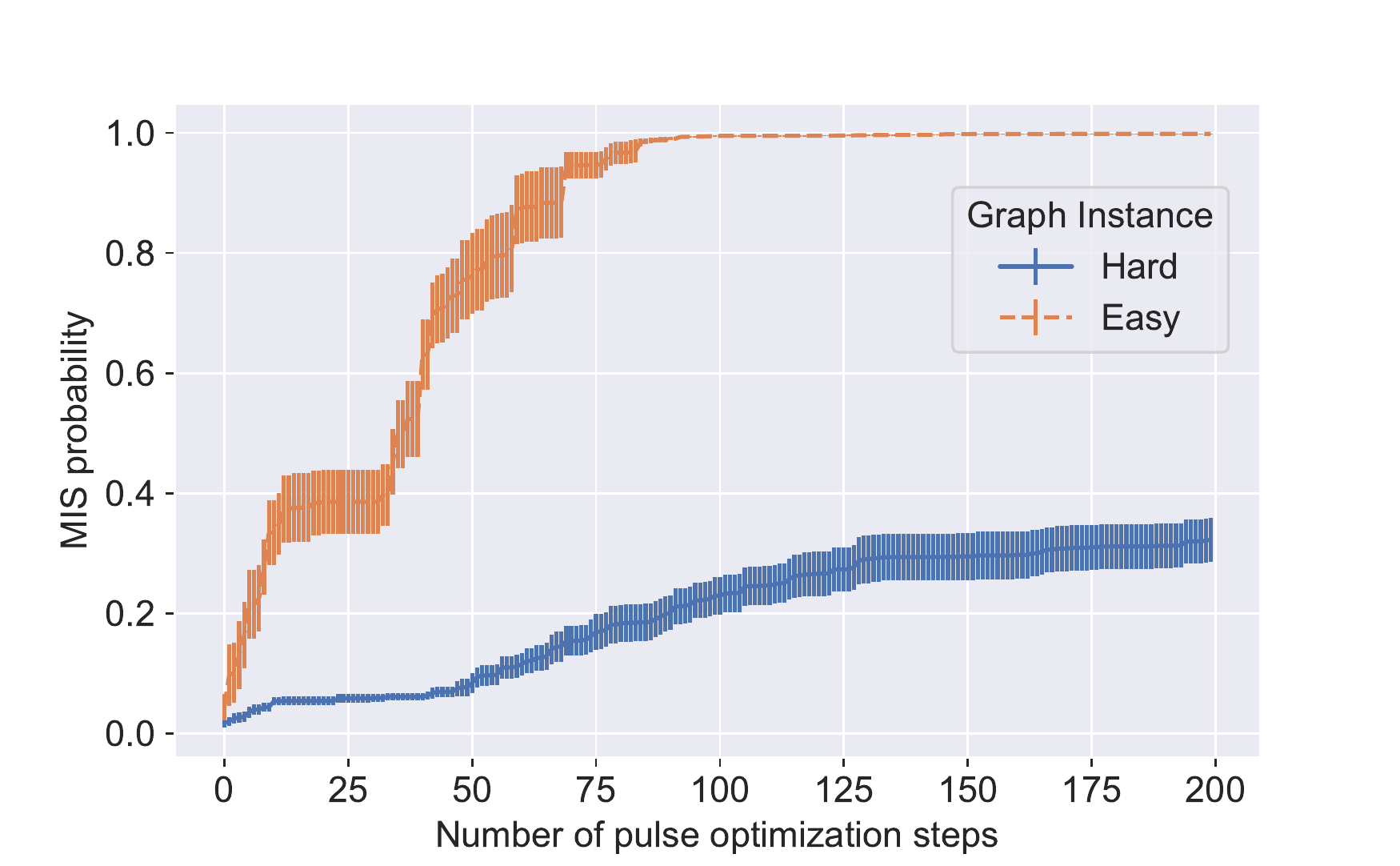}\captionsetup{justification=Justified}
\caption{Probability of reaching the MIS as a function of the number of pulse shaping steps for an easy and a hard graph instance (graphs depicted in Fig.~\ref{fig:easy_graph} and Fig.~ \ref{fig:hard_graph}, respectively).}
\label{fig:hard_vs_easy}
\end{figure}

For concreteness, consider the embedded graph of Fig.~\ref{fig:easy_graph}. It is a $3\times3$ square lattice, and therefore a UD graph. A pulse shaping routine can be written for finding the MIS of this graph, which is given by the node set $\{0,2,4,6,8\}$. The best shape resulting from a simple Bayesian search routine is shown in Fig.~\ref{fig:easy_pulse} and it is reminiscent of an adiabatic protocol. The system is initialized in $\ket{0} \otimes \ldots \otimes \ket{0}$, corresponding to the ground state of the Hamiltonian for small $\Omega$ and large negative detuning ($t \sim 0-100 ns$). $\Omega$ then ramps up and plateaus while the detuning slowly changes sign ($t \sim 100-1500ns$). Finally, $\Omega$ slowly ramps down again while the detuning keeps on increasing to large positive values ($t \sim 1500-3000ns$), where the excitations of the system correspond with a high probability to the MIS of the graph. The probability of measuring the MIS of the graph as a function of the number of Bayesian minimization iterations is shown in Fig.~\ref{fig:hard_vs_easy} under the label \textit{Easy graph instance}. For such an ideal graph with a clear UD embedding, the pulse shaping routine is able to reach on average a near-perfect MIS probability in less than a hundred optimization steps.

\begin{figure}[b]
    \centering\captionsetup{justification=Justified}
    \includegraphics[width=0.3\textwidth]{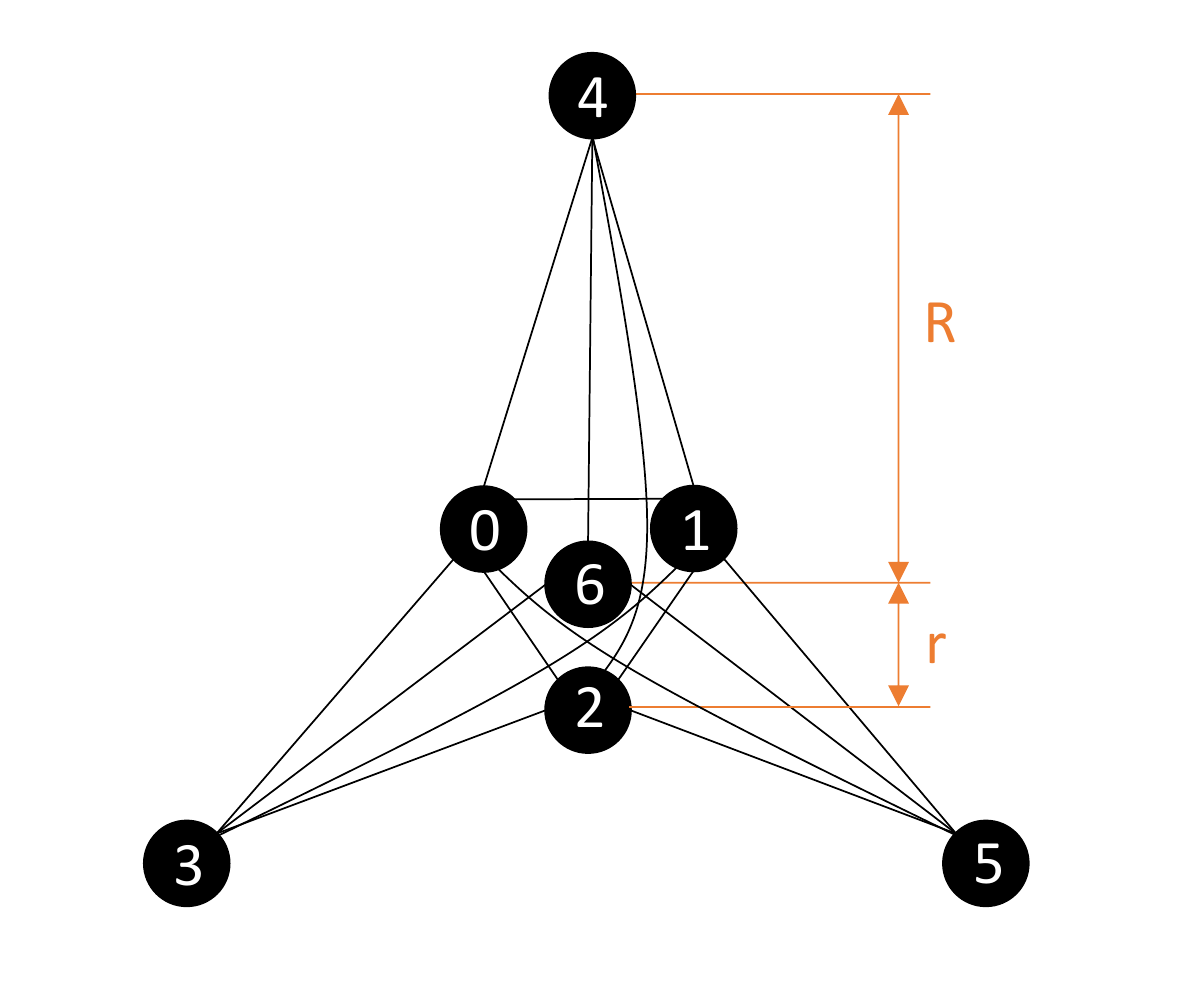}\captionsetup{justification=Justified}
    \caption{Different embedding for the same graph of Fig.~\ref{fig:hard_graph}. The parameters $R$ and $r$ indicate the size of the equilateral triangles formed by the nodes $\{3,4,5\}$ and $\{0,1,2\}$, respectively.}
    \label{fig:easier_graph}
\end{figure}

When a UD embedding is hard to find or does not exist, such a protocol becomes less effective. Consider the graph of Fig.~\ref{fig:hard_graph}, which is not a UD graph and whose MIS is given by the set of nodes $\{3,4,5\}$. Implementing pulse shaping naively without optimizing for the graph embedding yields substantially poorer results shown again in Fig.~\ref{fig:hard_vs_easy} under the label \textit{Hard graph instance}. The pulse shaping routine was not able to exceed a $35\%$ probability of finding the MIS after 200 optimization steps.

\begin{figure}[t]
    \centering
    \includegraphics[width=0.5\textwidth]{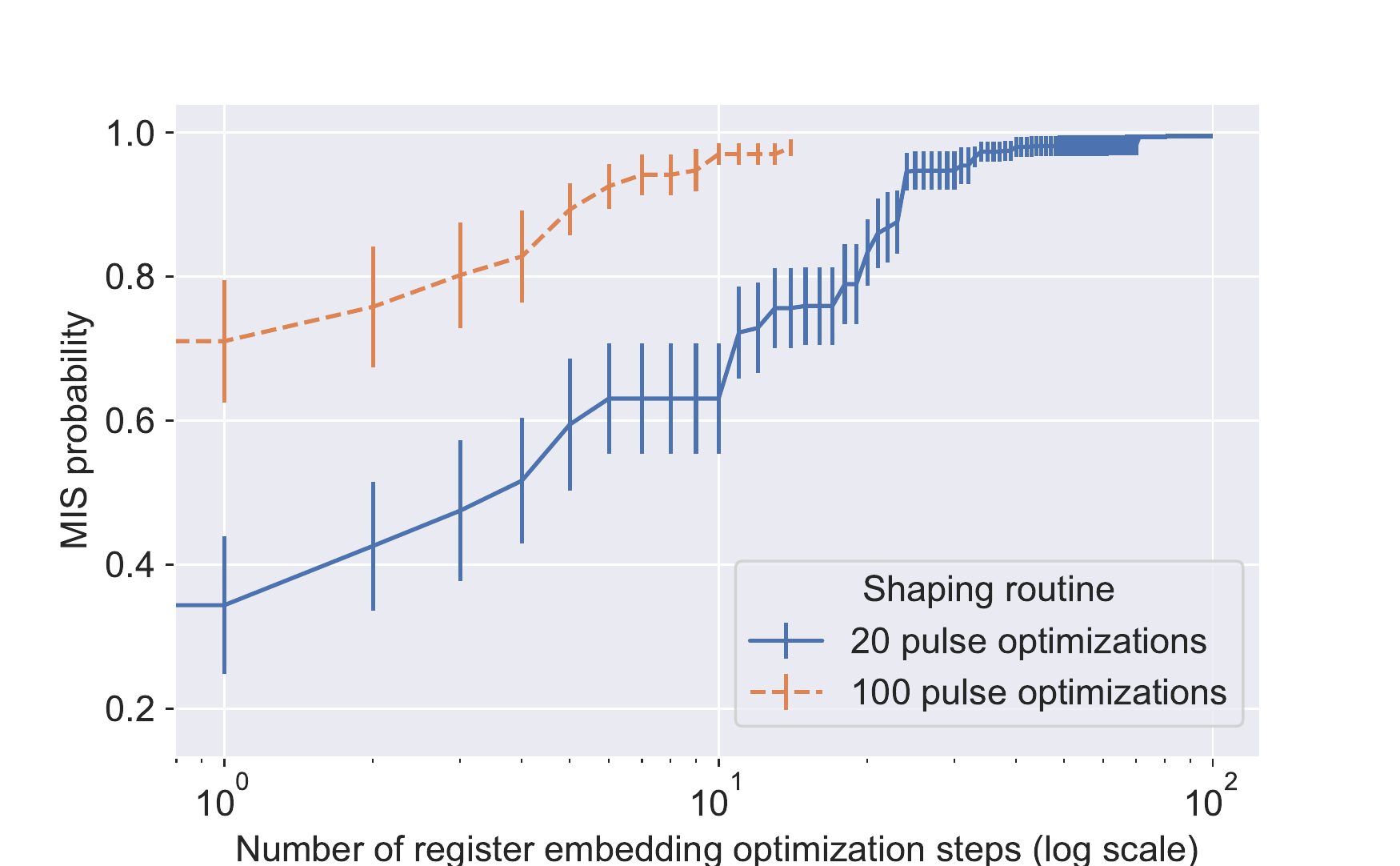}
    \captionsetup{justification=Justified}
    \caption{Probability of reaching the MIS as a function of the number of register embedding optimization steps as described in Fig.~\ref{fig:easier_graph}. For each register, a pulse shaping routine was called composed of 20 optimization steps (blue solid line) or 100 steps (orange dashed line).}
    \label{fig:easier_score}
\end{figure}
  
Choosing different embeddings for the graph can substantially change the probability of success of pulse shaping. For illustrative purposes, consider the same graph as Fig.~\ref{fig:hard_graph}, but with the embedding represented by Fig.~\ref{fig:easier_graph}, which depends on a choice of parameters $R$ (controlling the distance from the center of nodes 3, 4, and 5) and $r$ (controlling the distance from the center of nodes 0, 1 and 2). This embedding mimics and preserves some of the symmetries of the graph. Ideally, one wants $R > r$ so that the three nodes comprising the MIS are well separated and less subject to the Rydberg blockade mechanism. However, $R$ should still be small enough so that each of the qubits 3, 4, and 5 block the excitations of 0, 1, 2, and 6. 

If the pulse shaping routine is extended to include a search in the space of embeddings parametrized by $R$ and $r$, the probability of reaching the MIS increases substantially as shown in Fig.~\ref{fig:easier_score}. Two strategies are compared: one where 15 embedding steps are supplemented with 100 pulse shaping steps, and one where 100 embedding steps are supplemented with 20 pulse shaping steps. In this example, even a short optimization on the embedding can lead to a dramatic performance improvement compared to the naive embedding score. As illustrated in this example, if one wants to speed up the convergence of the optimization, one should balance the time spent between improving the embedding and the pulse shaping.

\subsection{Noise model}
The simple examples just presented assumed a perfectly noiseless machine. The current iteration of neutral atom devices, like any other digital or analog quantum device available at the time of writing, is not error corrected. The results are therefore expected to be limited by noise. In this respect, however, the analog mode of operation is more resilient (as it will be shown in Section~\ref{results}). A practical way of quantifying the noise on the machine is to send a constant laser pulse that would induce Rabi oscillations of known frequency onto single atoms, and measure the noise-induced damping of the oscillations~\cite{noisemodel}. Four types of error sources are identified:
\begin{enumerate}
    \item SPAM (state preparation and measurement) errors
    \item Non-zero temperature effects
    \item Imperfections in the driving laser
    \item Spontaneous emission 
\end{enumerate}

SPAM errors are parametrized by three probabilities $\eta$, $\epsilon$ and $\epsilon'$. The optical pumping process that prepares all the atoms in the $\ket{0}$ state might fail with probability $\eta$, resulting in some atoms not participating in the quantum evolution at all. 
The final state is measured by fluorescence imaging of the atom array, where atoms are ``bright" if in the ground state and ``dark" otherwise.
False positive and false negative probabilities, $\epsilon$ and $\epsilon'$, reflect the fact that some atoms in $\ket{0}$ might escape from the trap or leave the ground state due to collisions and be wrongly labeled as $\ket{1}$, while some atoms in $\ket{1}$ might decay in the ground state and be wrongly labeled as $\ket{0}$. Additionally, atoms are cooled to temperatures very close to absolute zero, of the order of $\mu$K. The residual thermal motion, however, induces a non-negligible Doppler shift in the detuning experienced by each atom. Another source of inhomogeneity comes from the fact that the driving lasers have a profile that is not perfectly flat, but rather more of a Gaussian shape. Therefore the atoms at the border of the register experience a lower amplitude than the ones at the center. Finally, modeling Rydberg atoms as two-level systems is a rather high-level description. The process of excitation from the ground state to the excited Rydberg state involves the transition to an intermediate state, from which the atom can spontaneously decay into some hyperfine ground state that will be eventually measured as $\ket{0}$ but can never be excited to $\ket{1}$ during the evolution. This process of spontaneous emission can be modeled as a dephasing channel in the density matrix formalism.

In what follows, we present new strategies to overcome the difficulties presented in this section.  
\section{Methods}
 \label{methods}
We dedicate this section to introducing our new approach to solving combinatorial graph problems using neutral atoms QPUs. First, we present different strategies to embed any class of graphs into atom registers. Moreover, in order to accelerate the pulse shape optimization process, we propose a supervised machine learning model capable of predicting pulses that are specifically tailored to find near-optimal solutions for a given combinatorial graph problem. Finally, we discuss how we assess the quality of the proposed quantum algorithm by calculating the related Q-score~\cite{qscore}. 

\subsection{Embedding strategies} 
\label{spring_section}
\begin{figure}[b]
\centering
\includegraphics[width=0.48\textwidth]{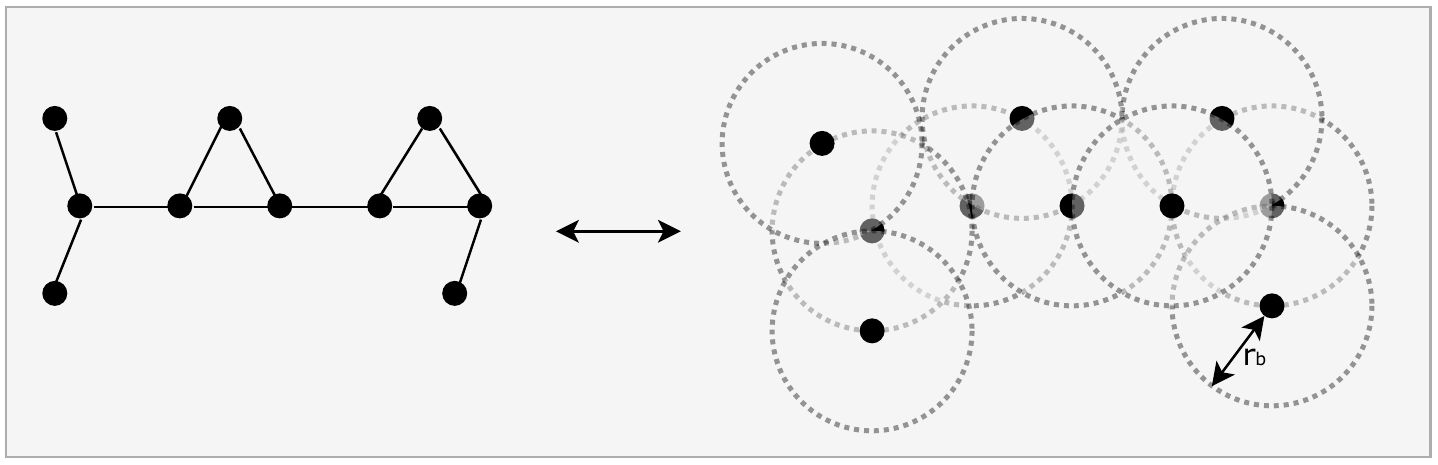}\captionsetup{justification=Justified}
\caption{Example of embedding a UD graph into a register, where two atoms strongly interact if they are within the blockade radius $r_b$ of each other.}
\label{ud_emb_ex}
\end{figure}

Unit-Disk (UD) graphs compose a special class that is naturally embedded in Rydberg atom-based QPUs. Any UD graph is composed of a set of nodes with their related positions in the Euclidean plane. For each pair $\{i,j\}$ of nodes, there is an edge connecting them if and only if their distance is below a fixed threshold $r$. As seen in Fig.~\ref{ud_emb_ex}, 2-dimensional Rydberg atom-based registers can be wisely created to match the graph under consideration. 

Setting the same blockade radius to all atoms, the resulting interactions can directly represent the connections on the related embedded graph. This implies that, due to the Rydberg blockade phenomenon, two connected atoms in $\ket{00}$ cannot be excited at the same time in $\ket{11}$, but instead they will form the entangled state $\ket{01} + \ket{10}$. Hence, any final state corresponds to an independent set in the related UD graph, and, by applying variational techniques, one can drive the system to find the largest set and hence solve the MIS problem.

However, finding a unit-disk realization for a given graph is proven to be NP-hard~\cite{breu1998unit} and, as the reader may anticipate, not all graphs have such an embedding (e.g., consider any $K_{1n}$ star graph with $n>6$). Furthermore, solving the MaxCut problem might potentially imply selecting two or more connected nodes, hence violating the blockade radius-based constraint. For instance, any cutting set of maximum size on complete graphs whose order is greater than three will have at least one pair of connected nodes. To overcome the aforementioned issues, other embedding strategies should be considered. For this purpose, we present here different embedding approaches based on the Fruchterman-Reingold algorithm~\cite{fruchterman1991graph}.

Force-directed algorithms are used to draw graphs in a plane in such a way that two connected (resp. disjoint) nodes are placed close to (resp. far from) each other, with a minimum (resp. maximum) distance between them (resp. from the plane's center). Fruchterman and Reingold also proposed in their work to place the vertices evenly in the frame and make the edges' lengths uniform in order to reflect inherent symmetries. For this purpose, each edge from the graph is treated as a spring that holds its endpoint vertices close to each other while a competing repulsive force is applied to push all vertices away from one another, even though they are not connected by an edge in the original graph. The iterations will stop when the system reaches the equilibrium, which minimizes the difference between all repulsive and attractive forces. 

The attractive and repulsive forces $f_a$ and $f_r$ between two nodes are respectively given by equations \eqref{attr} and \eqref{repul}, where $r_{ij}$ is the distance between the nodes $i,j \in \vertices$, while $k = \sqrt{area/|\vertices|}$ is set to be related to the area of the Euclidean plane. Moreover, the total energy $f_t$ of the system is given by adding the forces between all pairs of vertices, as shown in \eqref{total}. Hence, $f_t$ goes to zero as the system approaches its equilibrium (note that the repulsive forces cannot be positive). For a deep description of the algorithm, one may refer to ~\cite{fruchterman1991graph}. 
\begin{align}
   & f_a(i,j) = r_{ij}^2/k  \label{attr}\\
   & f_r(i,j) = -k^2/r_{ij} \label{repul}\\
    & f_t = \sum_{i,j \in \edges}f_a(i,j) + \sum_{i \in \vertices}\sum_{j \in \vertices: i\neq j}f_r(i,j) \label{total}
\end{align}

One advantage of this algorithm is that the graph under consideration is naturally embedded as a UD graph if such realization exists for it and if enough iterations are allowed (i.e., by iterating until the system reaches the equilibrium). Another interesting characteristic is that, by adding positive (resp. negative) weights to the graph's edges, one can give more (resp. less) importance to some specific pairs of nodes, hence placing them closer to (resp. further from) each other. Note, however, that such approaches might potentially return asymmetric topologies.

\begin{figure}[b] 
  \begin{subfigure}[b]{0.498\linewidth}
    \centering
    \includegraphics[width=0.55\linewidth]{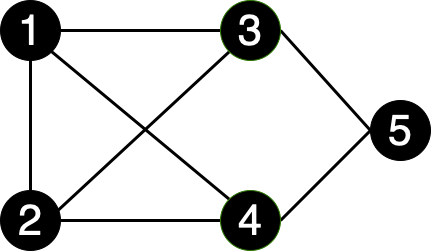} 
    \caption{Spring layout (SL)} 
    \label{sp} 
  \end{subfigure}
  \begin{subfigure}[b]{0.498\linewidth}
    \centering
    \includegraphics[width=0.55\linewidth]{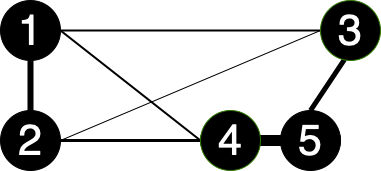} 
    \caption{Random weight SL} 
    \label{rawsp} 
  \end{subfigure} 
  \begin{subfigure}[b]{0.498\linewidth}
    \centering
    \includegraphics[width=0.55\linewidth]{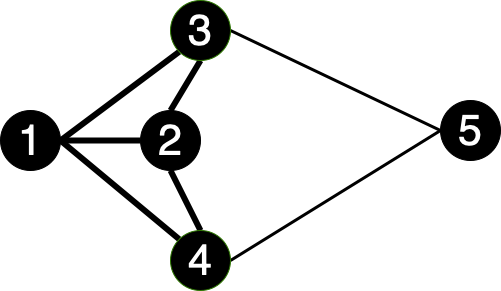} 
    \caption{Weighted spring layout} 
    \label{wesp} 
  \end{subfigure}
  \begin{subfigure}[b]{0.498\linewidth}
    \centering
    \includegraphics[width=0.55\linewidth]{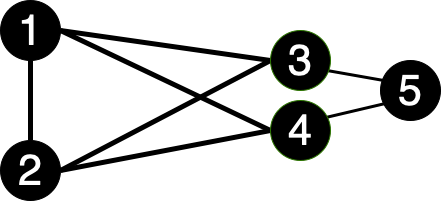} 
    \caption{Inverse-weight SL} 
    \label{wisp} 
  \end{subfigure} \captionsetup{justification=Justified}
  \caption{Illustration of different register layouts for the same graph instance: the positions were generated with the Fruchterman-Reingold algorithm. Edges with higher wights are represented with thicker lines.}
  \label{layouts} 
\end{figure}

Fig.~\ref{layouts} presents 4 different topologies based on the Fruchterman-Reingold algorithm on the same graph $G = (\vertices, \edges)$ with 5 nodes and 7 edges. While Fig.~\ref{sp} shows a possible embedding by directly applying the algorithm on $G$ (hereafter referred to as \textit{spring layout}), Fig.~\ref{rawsp} depicts a solution considering random values as edge weights. By applying such an approach, hereafter referred to as \textit{random weight spring layout}, adjacent nodes whose edges have higher weights (represented with thicker lines) are then placed closer to each other, hence breaking the layout symmetry. 

One might also want to give more importance to a sub-set of edges during the process of embedding. This can be done by setting specific weights to the corresponding edges. Here, we propose two different strategies that are specifically tailored to solve MaxCut and MIS instances.  Let $w(i,j)$ be the weight of the edge $\{i,j\} \in \edges$, and $N(i)$ be the number of neighbours of node $i \in \vertices$. Then, the weight of any edge $\{i,j\} \in \edges$ is set as the product of the related neighborhood size of its endpoints:  
\begin{align}
   & w(i,j) = N(i) N(j), & \forall \{i,j\} \in \edges \label{wsp_form}
\end{align}

The new related attractive $f_a$ is given by multiplying \eqref{attr} by $ w(i,j)$ and, as shown in Fig.~\ref{wesp}, the resulting embedding, named \textit{weighted spring layout}, creates clusters closer to nodes having the biggest neighborhoods. Finally, we propose the \textit{inverse-weight spring layout} (see Fig.~\ref{wisp}), where the edge's weight $w(i,j)$ is calculated as in \eqref{wsp_form} but multiplying the result by -1: 
\begin{align}
   & w(i,j) = -N(i) N(j), & \forall \{i,j\} \in \edges
\end{align}

 The initial position of each node might potentially be randomly generated, and the node to be embedded during each iteration might potentially be picked in a different order. Hence, the proposed algorithm might generate different outputs by running it several times. Moreover, one might desire to stop the algorithm before it converges to the system's equilibrium. This approach can be an interesting strategy to try several embeddings in a limited runtime. 
 
 It is worthwhile mentioning that all the proposed embedding strategies are feasible on neutral atom-based QPUs once they respect the device's technical constraints, such as minimum distance between atoms and maximum distance from the register's center. If either technical constraint is violated, one might try re-scaling every position vector by a factor $\alpha > 0$. Sometimes, however, a specific embedding cannot satisfy both constraints at once. In that case, a different embedding strategy must be employed.

\subsection{Chained multi-target regression algorithm}
As seen in the previous sections, the way combinatorial graph problems are usually solved with quantum hardware involves the optimal tuning of a set of parameters. This is usually done via an optimization loop that is applied to each instance of the problem, which is time and resource-consuming. To overcome both time and resource limitations, we propose a new supervised machine learning-based approach that automates the parameter choices and creates pulse sequences for analog quantum processes. By predicting essential pulse parameters, one can considerably scale up quantum algorithms and, hence, solve bigger instances of complex combinatorial problems without dedicated optimization loops. 

To the best of our knowledge, only two machine learning techniques were proposed in order to accelerate Quantum Approximate Optimization Algorithms (QAOAs). To solve combinatorial problems, Khairy {\sl et al}~\cite{Khairy_2020, khairy2019reinforcement} propose two different machine learning-based approaches to find optimal QAOA parameters: a kernel density estimator-based model~\cite{Khairy_2020} that learns generative models of optimal circuit parameters, and a reinforcement learning-based model~\cite{ khairy2019reinforcement} that can learn different policies to predict (near-)optimal QAOA parameters. Comparing both proposed approaches and the optimization loop under limited runtime constraints, the authors showed that the optimality gap could be considerably reduced. Even though different machine learning-based approaches were proposed in order to find near-optimal parameters for circuit-based QAOA algorithms, no attention has been given to analog quantum processing on neutral-atom QPUs.

The main objective of our supervised machine learning-based approach is to automatically provide: i) the Rabi frequency and detuning values on different instants of the pulse, and ii) the total duration of the pulse. Hence, the out-coming pulse is specifically tailored to evolve the system to states that represent (near-)optimal solutions for a given combinatorial graph problem instance. In what follows, we detail each step of the proposed machine learning algorithm.

\subsubsection{Generating the training data set}\label{sec:tradata}
One of the most important steps of training a supervised machine learning model (SMLM) is the generation of a representative training data set (TDS) from which the model will learn what is a good solution for a given instance of the problem. For instance, for training an SMLM to predict pulse shapes for unseen instances, the TDS must be composed of good solutions for different instances of the same problem. Also, the TDS must cover a representative part of the possible input space. For graph problems, for example, the TDS should provide good pulse shapes for a set of heterogeneous instances that vary in order (number of nodes), size (number of edges), and register topology (atoms' position).

One effective way to generate such a TDS is applying the pulse optimization for each generated instance: for a given graph instance, one should find a near-optimal pulse shape for different register topologies. As presented in the previous sections, the closed optimization loop can be done by applying Stochastic Gradient Descent~\cite{mason1999boosting} or Bayesian optimization~\cite{frazier2018bayesian} algorithms on the pulse parameters for example. 

As a result, the TDS will provide two categories of data, named \textit{input features} and \textit{target values}. While the former provides some graph- and register-specific features, the latter contains the value for each parameter of the pulse. In what follows, we present examples of information fed to the SMLM during the training process:

\begin{itemize}
    \item \textit{Input Features}: graph order (number of nodes), graph size (number of edges), graph density, minimum/maximum/average neighborhood size, minimum/maximum/average distance (in $\mu$m) between connected and disjoint nodes in the related register, and number of pulse points to be predicted. 
    \item \textit{Target Values}: Rabi frequency and detuning values (in rad/$\mu$s) to each point of the pulse and its total duration~(in~$\mu$s).
\end{itemize}

It is worthwhile mentioning that the points of the pulse are evenly spaced related to the whole duration of the waveform. Also, note that the input features do not depend on the size of the instance as they are present in any graph and register.

\subsubsection{Learning method}\label{sec:cmlearn}
We propose a supervised machine learning approach based on the Chained Multi-Target Regression Algorithm (CMTRA)~\cite{demirel2019ensemble}. This method is generally used to predict multiple target values that are dependent upon the input and upon each other. The CMTRA can be formally defined as follows. Let $\mathcal{D}$ be a training data set with \textit{N} instances. Each instance $n \in \{1,..,N\}$ is composed of a vector of input features $x^n = (x^n_1, ..., x^n_{|x^n|})$ and a target vector $y^n = (y^n_1, ..., y^n_{|y^n|})$. Learning to predict $y$ from an input $x$ consists in finding a parameter estimator model $f$ that assigns values for each element of $y^n$ from a given instance $n$ and its related input vector $x^n$:
\begin{align}
    f(n) : x^n = (x^n_1, ..., x^n_{|x^n|}) \mapsto y^n = (y^n_1, ..., y^n_{|y^n|})
\end{align}
Now, let $X$ and $Y$ be respectively the set of all input feature vectors $x^n$ and target value vector $y^n$ from the decomposition of $\mathcal{D}$. Also, let $Y_j$ be the column vector for each target value type $j \in \{1,..,|y^n|\}$ whose  elements represent a specific target value related to each instance~$n \in N$. 

\begin{figure}[t] 
  \begin{subfigure}[b]{1\linewidth}
    \centering
    \includegraphics[width=.6\linewidth]{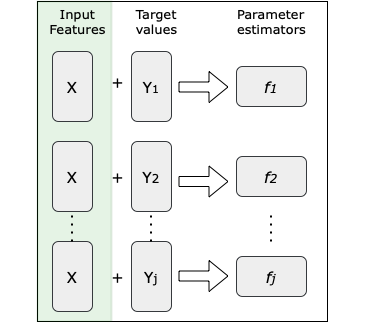} 
    \caption{Classical Multi-Target Regression algorithm.} 
    \label{mtr} 
  \end{subfigure}
  \\
  \begin{subfigure}[b]{1\linewidth}
    \centering
    \includegraphics[width=.6\linewidth]{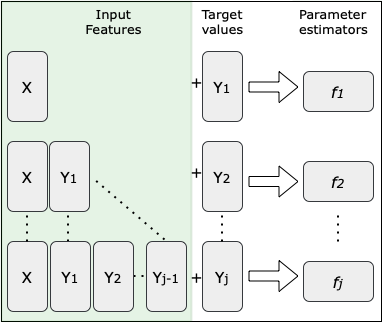} 
    \caption{Chained Multi-Target Regression algorithm.} 
    \label{cmtra} 
  \end{subfigure} \captionsetup{justification=Justified}
  \caption{Multi-Target Regression and Chained Multi-Target Regression algorithms.}
  \label{MTRA} 
\end{figure}

As shown in Fig.~\ref{mtr}, classical multi-target regression algorithms generally learn a specific inner parameter estimator model $f_j$ for predicting each target value $y^n_j \in Y$ separately. The learning process is done $|y^n|$ times (once for each target value) and takes only the feature vectors $x^n \in X$ as input. With CMTRA, however, for each new parameter estimator $f_j$, the $Y_{j-1}$ target vector is added to the original input feature vectors $X$ to be used during the current learning process (see Fig.~\ref{cmtra}).

As a parameter estimator model, we use the Gradient Boosting Regressor (GBR)~\cite{friedman2001greedy} on each learning iteration throughout the prediction chain. GBR is based on decision trees built in a stage-wise manner with several decision trees as weak learners that, combined, provide better predictions. The algorithm is implemented in such a way to minimize a given loss function, which is related to the gap between the predicted values $\hat{y}_i$ and the expected ones in $y_i$ given an input vector $x$. In this work, we apply the Squared Error Loss (SEL) function to qualify the prediction of any target value. The SEL function can be defined as follows:  
\begin{align}
     SEL(j) =  \sum_{n = 1}^N (\hat{y}_j^n - y_j^n)^2 & \forall j \in \{1,..,|y^n|\} 
\end{align}

For an in-depth description of Chained Multi-Target Regression and Gradient Boosting Regressor algorithms, one may respectively refer to~\cite{demirel2019ensemble} and~\cite{friedman2001greedy}.


\subsection{Q-score metric }

In recent years several proposals have been put forward to assess the performance of quantum computers. Early protocols such as Randomized Benchmarking~\cite{RB2011, RB2012} and Quantum Process Tomography~\cite{QPT2013} aim at benchmarking the fidelity of quantum gates and circuits without a specific application in mind, and an analog version had to be developed separately~\cite{Shaffer2021}. The more recent proposal of Local Hamiltonian Learning~\cite{LHL2019} can be adapted to both the digital and analog paradigm~\cite{HT2020}, but again with no focus on the actual problem-solving capabilities of the device. The Q-score metric~\cite{qscore} was developed to overcome these limitations at a time when commercially viable NISQ applications are becoming a reality. It is application-centric, hardware-agnostic, and can be applied equally effectively on current machines as well as future large-scale devices. For these reasons, the Q-score represents to date one of the best attempts at establishing a practical standardized benchmark that can be monitored over time to assess the evolution of quantum computers in solving real problems.

Essentially, the Q-score is comprised of the following steps:
\begin{enumerate}
    \item Pick a hard combinatorial optimization problem $P$ with input size $n$
    \item Establish the scaling in $n$ of the score of an optimal solution $\text{Opt}(n)$ and a random solution $\text{Rand}(n)$
    \item Solve several instances of $P$ on a quantum computer and calculate the average quantum score $\text{Quant}(n)$
    \item Calculate the improved approximation ratio $\beta(n)$:
    \begin{equation}\label{eq:impapprat}
    \beta(n) := \frac{\text{Quant}(n)-\text{Rand}(n)}{\text{Opt}(n)-\text{Rand}(n)}
    \end{equation}
    \item The Q-score is the largest integer $n^*$ such that $\beta(n^*) > 0.2$, meaning the largest problem size for which a quantum algorithm outperforms a random algorithm by at least 20\% \footnote{This arbitrary threshold was chosen by the authors of the original paper and for coherence, the same value will be used here.}.
\end{enumerate}

\begin{figure}[t] 
    \begin{subfigure}[b]{.49\linewidth}
    \centering
    \captionsetup{width=1\linewidth}
    \includegraphics[width=\linewidth]{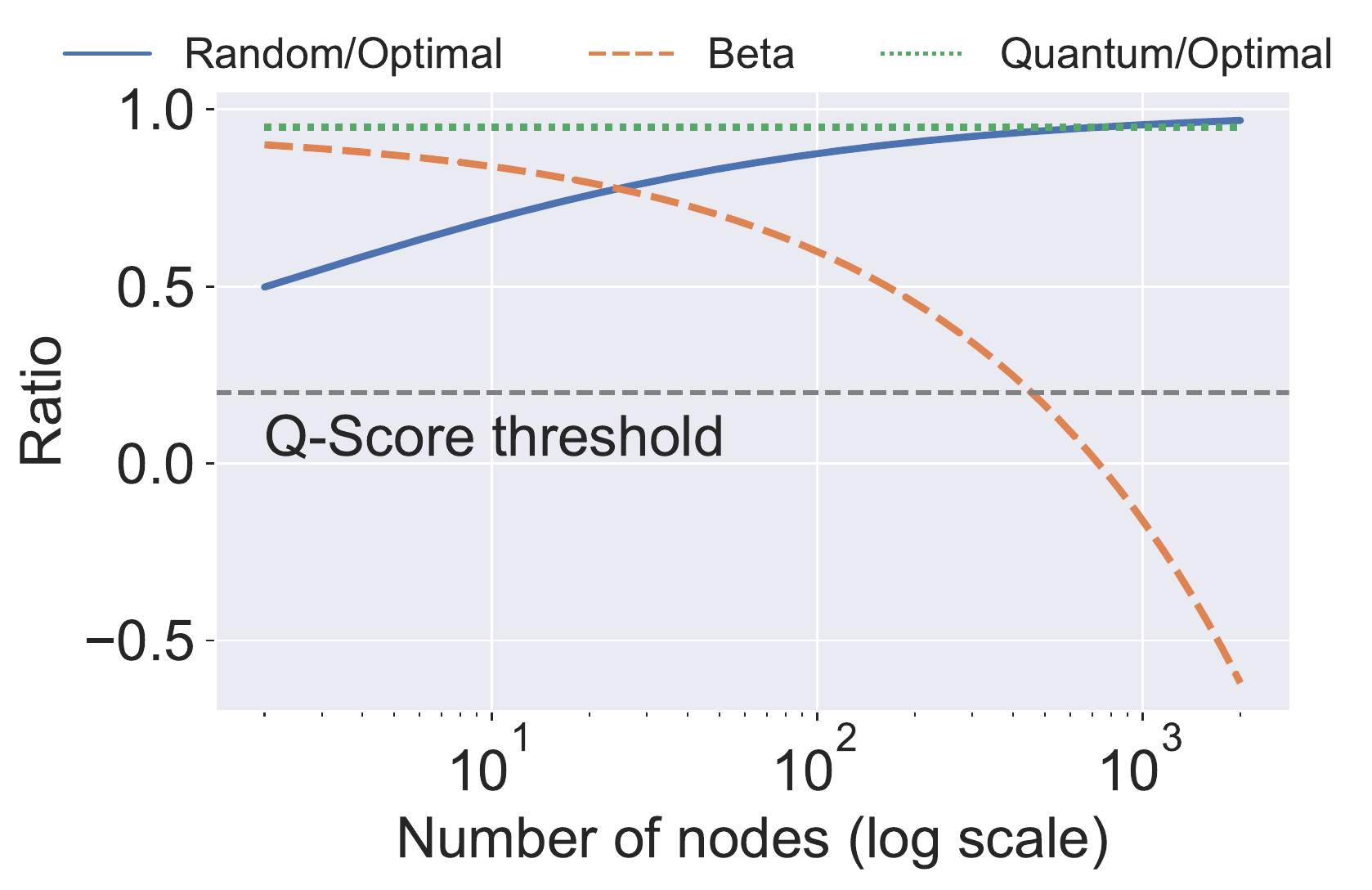} 
    \caption{Hypothetical algorithm.}
    \label{fig:beta_far}
  \end{subfigure}
  \begin{subfigure}[b]{.49\linewidth}
    \centering 
    \captionsetup{width=1\linewidth}
    \includegraphics[width=\linewidth]{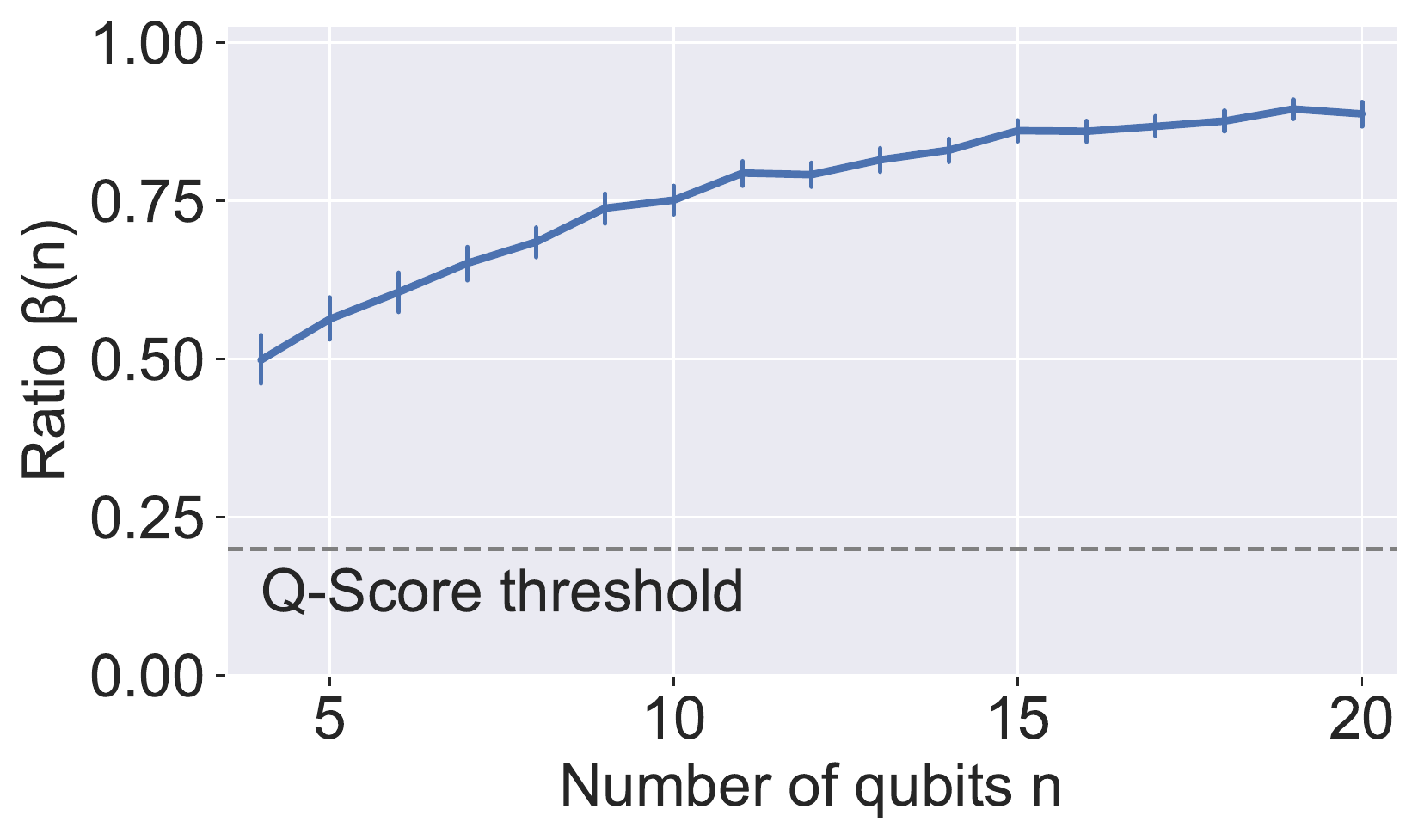} 
        \caption{Exact algorithm.}
    \label{fig:beta_fse}
  \end{subfigure}\captionsetup{justification=Justified}
  \caption{
  \ref{fig:beta_far} The $\beta$ function for MaxCut on $\mathcal{G}(n,0.5)$ (orange dashed line) of a hypothetical algorithm with fixed approximation ratio $\text{Quant}(n) = 0.95 \text{Opt}(n)$ (green dotted line) will inevitably decrease due to $\text{Rand}(n)/\text{Opt}(n)$ (blue line) being asymptotically saturated.
  \ref{fig:beta_fse} $\beta$ function of an exact algorithm for MaxCut on random $\mathcal{G}(n,0.5)$ graphs. A value of 1 is only reached asymptotically, where \eqref{eq:optn} and \eqref{eq:randn} are respectively chosen as expected optimal and random values.}
\end{figure}

These steps describe a slightly more general framework than the original definition of the Q-score metric. The Q-score was originally proposed for the MaxCut problem on the class of Erd\"os-Renyi graphs with an edge probability of 0.5, denoted here $\mathcal{G}(n, 0.5)$, because of the existence of rigorous scaling bounds on the value of random and optimal cuts~\cite{exact_scalings}: 
\begin{align}
    \text{Opt}(n) &= \frac{n^2}{8} + \lambda n^{\frac{3}{2}} \label{eq:optn} \\
    \text{Rand}(n) &= \frac{n^2}{8} \label{eq:randn}
\end{align}
with $\lambda\approx 0.178$. An infinite Q-score is possible in this case as long as the algorithm scales as $\text{Rand}(n) + \mu n^{3/2}$ with $0.2\lambda<\mu \le \lambda$. Algorithms with a fixed approximation ratio, {\sl i.e}. $\text{Quant}(n) = \alpha \text{Opt}(n)$, will inevitably pass the 0.2 threshold and even go to negative values of $\beta$, as shown in Fig.~\ref{fig:beta_far} with $\alpha$ set to 0.95. 

Typically, a noisy quantum algorithm is expected to reduce to random sampling for large enough problem instances, hence becoming no better than random. From this example, one can understand the choice of adjusting the na\"ive approximation ratio
$$ \frac{\text{Quant}(n)}{\text{Opt}(n)} $$
by subtracting from the numerator and denominator the random part of the score scaling as in \eqref{eq:impapprat}: the asymptotic contribution being the same for an optimal and a randomized algorithm, the approximation ratio can be artificially increased for any algorithm that performs at least as good as random by simply increasing the size of the problem. The $\beta$ ratio \eqref{eq:impapprat}, on the other hand, avoids such situations by being identically zero on random algorithms.

Lastly, it is worth noting that the use of asymptotic formulas for $\text{Opt}(n)$ and $\text{Rand}(n)$ is not necessarily the best option for all $n$, as finite-size effects might alter the results for small systems. For MaxCut on $\mathcal{G}(n, 0.5)$, for example, Eq. \eqref{eq:optn} overestimates slightly the average MaxCut value of small graphs, so that, as shown in Fig.~\ref{fig:beta_fse}, replacing $\text{Quant}(n)$ with the expected MaxCut value returned by an exact algorithm\footnote{An algorithm that always returns an optimal solution. The solutions were found by a classical solver as previously discussed.} on real instances
$$ \text{Quant}(n) \rightarrow \mathbb{E}
\left[ \text{MaxCut}(G) \right]_{G \in \mathcal{G}(n, 0.5)} $$
yields $\beta(n) < 1$. Therefore, a better choice for small system sizes would be to replace $\text{Opt}(n)$ and $\text{Rand}(n)$ with numerical values obtained empirically. Another reason to calculate $\text{Opt}(n)$ and $\text{Rand}(n)$ numerically is to extend the scope of Q-score to different graph classes or different combinatorial optimization problems for which rigorous asymptotic scalings are not known.


\section{Numerical Simulations}
 \label{results}
We now present the results of the numerical simulations. We detail first the emulation setup for generating the training data set, as well for predicting the sequences with the proposed machine learning-based approach.

\subsection{Simulation setup}
Let us first describe the setup used in our numerical simulations. While random graph instances were generated with the Vladimir-Brandes algorithm~\cite{batagelj2005efficient}, which produces Erd\H{o}s–R\'{e}nyi graphs, UD graphs were produced as proposed in~\cite{penrose2003random}. For each graph, we set the probability $p$ of connecting any pair of vertices with an edge to 0.5. It is worthwhile to mention that random graphs with the aforementioned density are unlikely to be unit-disks. Indeed, the probability of having a UD Erd\H{o}s–R\'{e}nyi graph quickly approaches zero as the number of the nodes increases and the density remains stable at 50\%. For this reason, this graph class ({\sl i.e.}, random graphs) is hereafter referred to as non-UD graphs.  The optimal solution of each instance was found by exactly solving the related Integer Linear Programming formulations for both MaxCut and MIS problems \cite{rendl2010solving,xiao2017exact}. Finally, random solutions were calculated as the average cost over one thousand random partitions of the vertex set.

All numerical simulations of the quantum device were designed using \textit{Pulser} \cite{silverio2022pulser}, an open-source python library for programming neutral-atom devices at the pulse level with high fidelity. In Pulser, a pulse can be built by specifying two time-dependent waveforms: one for the Rabi frequency of the laser and one for the detuning. Each waveform was obtained by interpolating between five free points equally spaced along the pulse duration. Fixing the initial and final value of the Rabi frequency to zero gives a total of nine free parameters: three for $\Omega$, five for $\Delta$, and one for the pulse duration. The resulting waveforms are of the type shown in Fig.~\ref{fig:easy_pulse}, where the round markers correspond to the points between which the curve is interpolated. The parameters were bound by realistic hardware specifications.

In order to find (near-)optimal pulse parameters, the closed-loop optimization was done by applying the Gradient Boosted Regression Tree algorithm from \textit{Scikit-Optimize} package~\cite{skopt} on the function \eqref{eq:avgcost}. We respectively set the maximum number of random starts ({\sl i.e.}, random guesses) and calls to the cost function to 10$n$ and 50$n$, where $n$ is the number of atoms in the related register. Let us recall that the cost function takes pulse's parameters as input variables to run the related sequence, and returns the solutions of the current instance. It is also worth mentioning that, due to resource limitations, the number of solved graphs exponentially decreases as the number of qubits increases: for each embedding strategy, 500 (resp. 10) graphs with 6 (resp. 16) nodes were solved for both MaxCut and MIS problems, on average. 

Given a graph instance, we applied the described closed-loop optimization process on four different registers, which were created by applying the embedding strategies presented in Section~\ref{spring_section}. To this end, each atom's position was found with the \textit{spring layout} function from \textit{Networkx} package~\cite{hagberg2008exploring}: while the edges' weights were set as previously described, the maximum number of iterations was limited to 100. Moreover, we multiplied each position vector by 40 in order to respect the distance constraints imposed by the device. The register and final pulse shape ({\sl i.e.}, after the closed-loop optimization process) that maximized the size of the independent (resp. cut) set among those sampled in 1000 runs were then selected as the final solution. Outputs from the closed-loop optimization ({\sl i.e.}, registers and pulse sequences) were then saved as training instances along with the related graph. From each solved instance, the data used as input features to train the ML model were those presented in Section~\ref{sec:tradata}. Finally, the TDS was generated only with Erd\H{o}s–R\'{e}nyi graphs, which are not necessarily unit-disks, and on a noiseless setup.

We implemented the CMTRA in python language using the \textit{Sklearn} package~\cite{sklearn}, where all parameters were set as default. The model training was done by applying the CMTRA with the generated training data set as discussed in Section~\ref{sec:cmlearn}. Once trained, the SMLM was applied to predict pulses for new, unseen graph instances for both MaxCut and MIS problems. Moreover, we generated 10 different registers for each graph instance, whose topologies were randomly selected among those presented in Section~\ref{spring_section}: the final register topology and the predicted pulse were chosen as previously discussed.  
\begin{table}[t]\footnotesize
  \centering
    \begin{tabular}{|c|c|c|}\hline
    \multirow{3}[1]{*}{~~SPAM~~} & $\eta$   & 0.005 \\
\cline{2-3}          & $\epsilon$   & 0.03 \\
\cline{2-3}          & $\epsilon'$  & 0.08 \\\hline
    \multicolumn{2}{|c|}{~~Temperature~~} & $30 \mu K$ \\\hline
    \multicolumn{2}{|c|}{Laser waist} & ~~$148 \mu m$~~ \\\hline
    \end{tabular}
  \caption{Noise parameters used in \textit{noisy}$^+$ emulations.}
  \label{tab:noise}
\end{table}

The evolution of the quantum system under the predicted pulse for a given register was then simulated using Pulser's simulation module, which in turn relies on the \textit{QuTiP} package~\cite{qutip}. Both noiseless and noisy simulations were performed. Noiseless simulations involve solving the time-dependent Schr\"{o}dinger equation. The output of a noiseless simulation is a vector in the Hilbert space that can be sampled a finite number of times in order to mimic a real experimental setup with a limited measurement budget. Noisy simulations, on the other hand, can be rather cumbersome depending on the type of noise to be included. The least expensive noise source is related to measurement errors $\epsilon$ and $\epsilon'$, and can be calculated by post-processing any kind of state sampling. Including preparation errors, laser defects and temperature effects would require, in principle, performing a new simulation for each sample that has to be collected. To find a compromise between computational resources and simulation accuracy, we decided to perform only five independent noisy simulations and to collect 100 samples from each. The last noise type available in Pulser, the dephasing channel accounting for spontaneous emission, would force the adoption of the density matrix formalism and the solution of the Lindblad master equation~\cite{manzano2020short}, introducing a rather severe computational overhead. For this reason, and because it is expected to have a smaller effect for short pulses, noisy simulations included in this work did not take dephasing into account. Two sets of noisy emulations were then performed. The first, denoted \textit{nosiy}$^+$, uses the noise parameters summarized in Table \ref{tab:noise} and is based on current hardware specifications. The second, denoted \textit{noisy}$^-$, has SPAM error rates halved.

\subsection{Results}
\begin{figure}[t] 
    \begin{subfigure}[b]{1\linewidth}
    \centering
    \captionsetup{width=1\linewidth}
    \includegraphics[width=\linewidth]{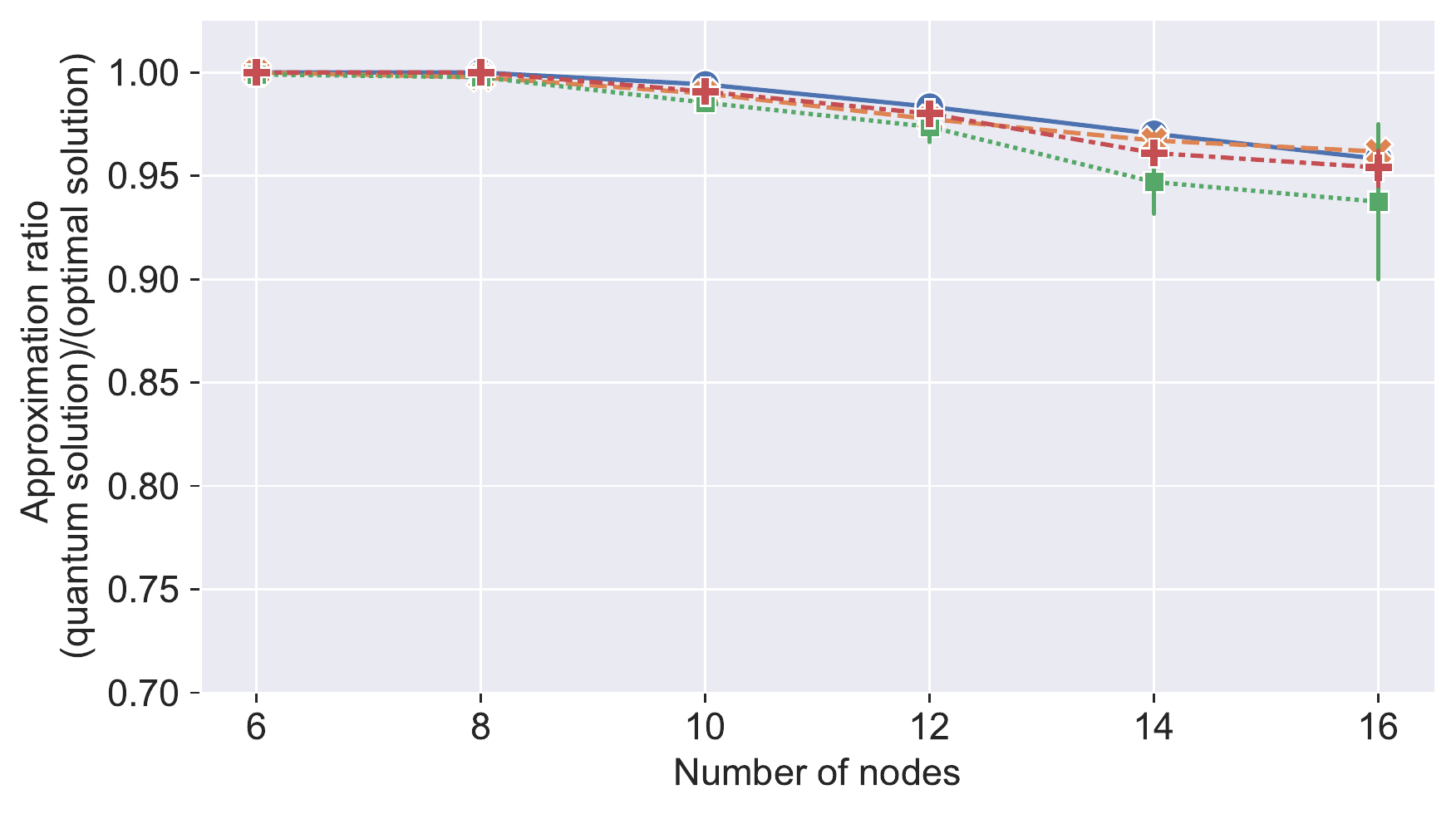} 
    \caption{Solution quality for MaxCut on non-UD graphs.}
    \label{fig:betagfhf_far}
  \end{subfigure}
  \begin{subfigure}[b]{1\linewidth}
    \centering 
    \captionsetup{width=1\linewidth}
    \includegraphics[width=\linewidth]{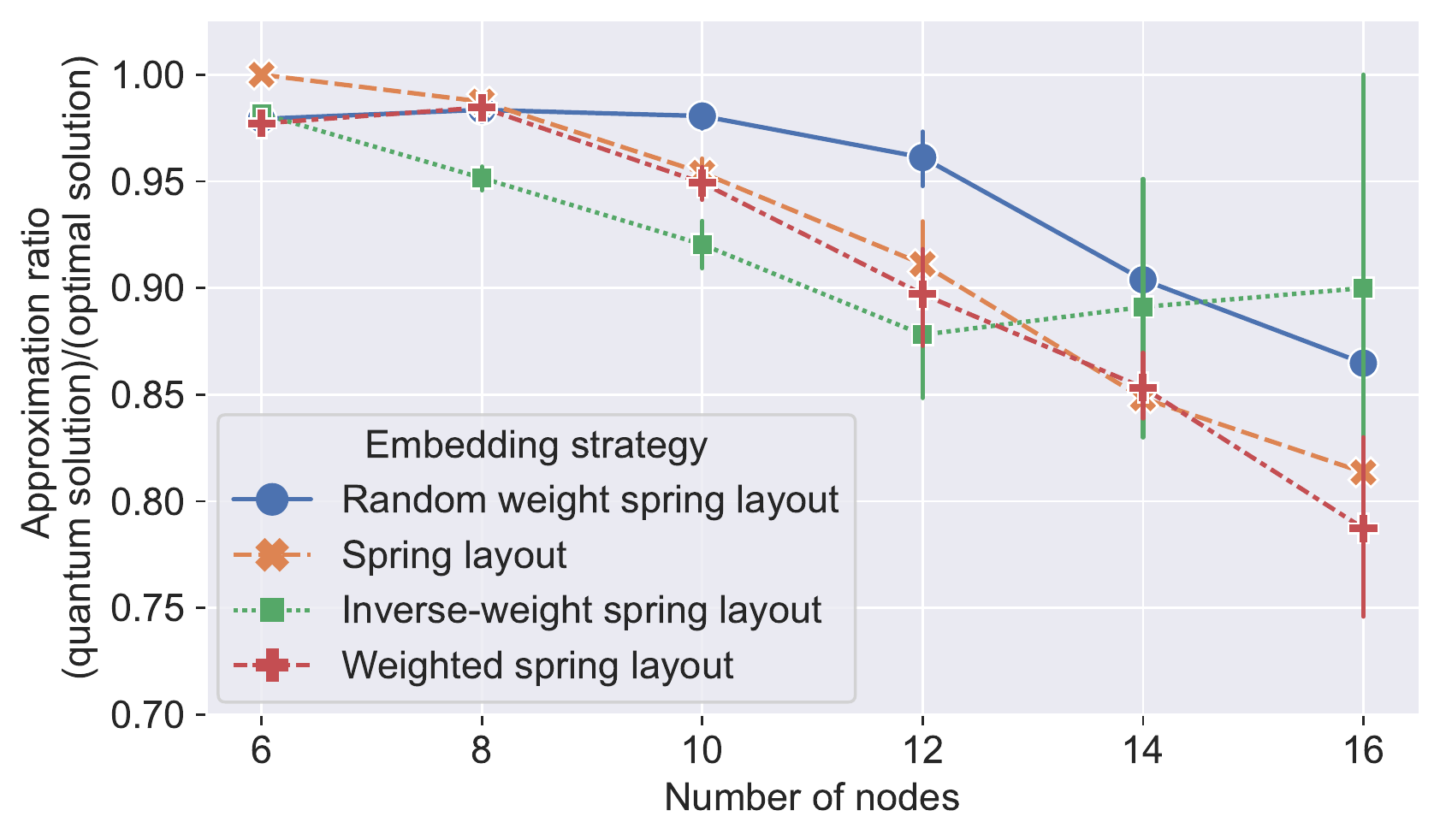} 
          \caption{Solution quality for MIS on non-UD graphs.}
    \label{fig:betfaghg_fse}
  \end{subfigure}
    \captionsetup{justification=Justified}
  \caption{Training data set generation: evolution of approximation ratio using closed-loop optimization on different embedding strategies and combinatorial graph problems.}
  \label{fig:results_closed_loop}
\end{figure}

Let us first present the quality of the generated training data set. Fig.~\ref{fig:results_closed_loop} depicts the evolution of the approximation ratio ({\sl i.e.}, the ratio of the cost of the quantum solution over the optimal one) on the generated TDS using closed-loop optimization on different embedding strategies and combinatorial graph problems. For both problems and each register size (graph order), we show the mean and the standard variation of the approximation ratio with a 0.95 confidence interval. It is worth mentioning that all TDS was generated on a noiseless setup and with Erd\H{o}s–R\'{e}nyi graphs, which are, with high probability, not UD graphs.

As seen in Fig.~\ref{fig:results_closed_loop}, neutral atom-based QPUs can efficiently solve both unconstrained and constrained classes of combinatorial graph problems, even on non-UD graphs. For instance, by optimizing the pulse shape for each instance of the MaxCut problem, the best solution was always found for 6 and 8-node graphs (see Fig~\ref{fig:betagfhf_far}). Moreover, while MaxCut instances are not strongly impacted by the embedding strategy, we observed that MIS instances could be better solved with random weight layout in general: compared to weighted spring layout (resp. spring layout), the average approximation ratio on 12-node (resp. 14-node) graphs could be improved by roughly 6\% (see Fig~\ref{fig:betfaghg_fse}). Also, layouts that put high-degree adjacent nodes far from each other appear to have a positive impact on bigger instances: compared to weighted spring layout (resp. spring layout), the average approximation ratio on 16-node  graphs could be improved by roughly 14\% when inverse weight spring layout was applied (see Fig~\ref{fig:betfaghg_fse}).

As also observed in Fig.~\ref{fig:results_closed_loop}, the quality of the solutions for MaxCut is better than for MIS, especially on small graphs. This behavior can be partially explained by the fact that, unlike MIS, MaxCut is an unconstrained problem, and hence any solution is a feasible solution for a given instance of the problem. This feature creates several symmetric solution subsets ({\sl i.e.}, subsets of different solutions with the same cut size), which facilitates the search for an optimal cutting set. However, increasing the maximum number of random starts and calls to the cost function during the optimization process, {\sl e.g.}, might potentially increase the overall quality of MIS solutions. 

In what follows, we present the results for the MaxCut and MIS problems with the proposed machine learning-based CMTRA model as a pulse predictor. Let us recall that no closed-loop optimization was done to find good pulse shapes since the CMTRA model was previously trained to predict the related parameters only taking graph and register features as input, as previously discussed.\vspace{-0.14cm}

\subsubsection{MaxCut problem}

\begin{figure*}[t] 
    \begin{subfigure}[b]{0.49\linewidth}
    \centering
    \captionsetup{width=1\linewidth}
    \includegraphics[width=\linewidth]{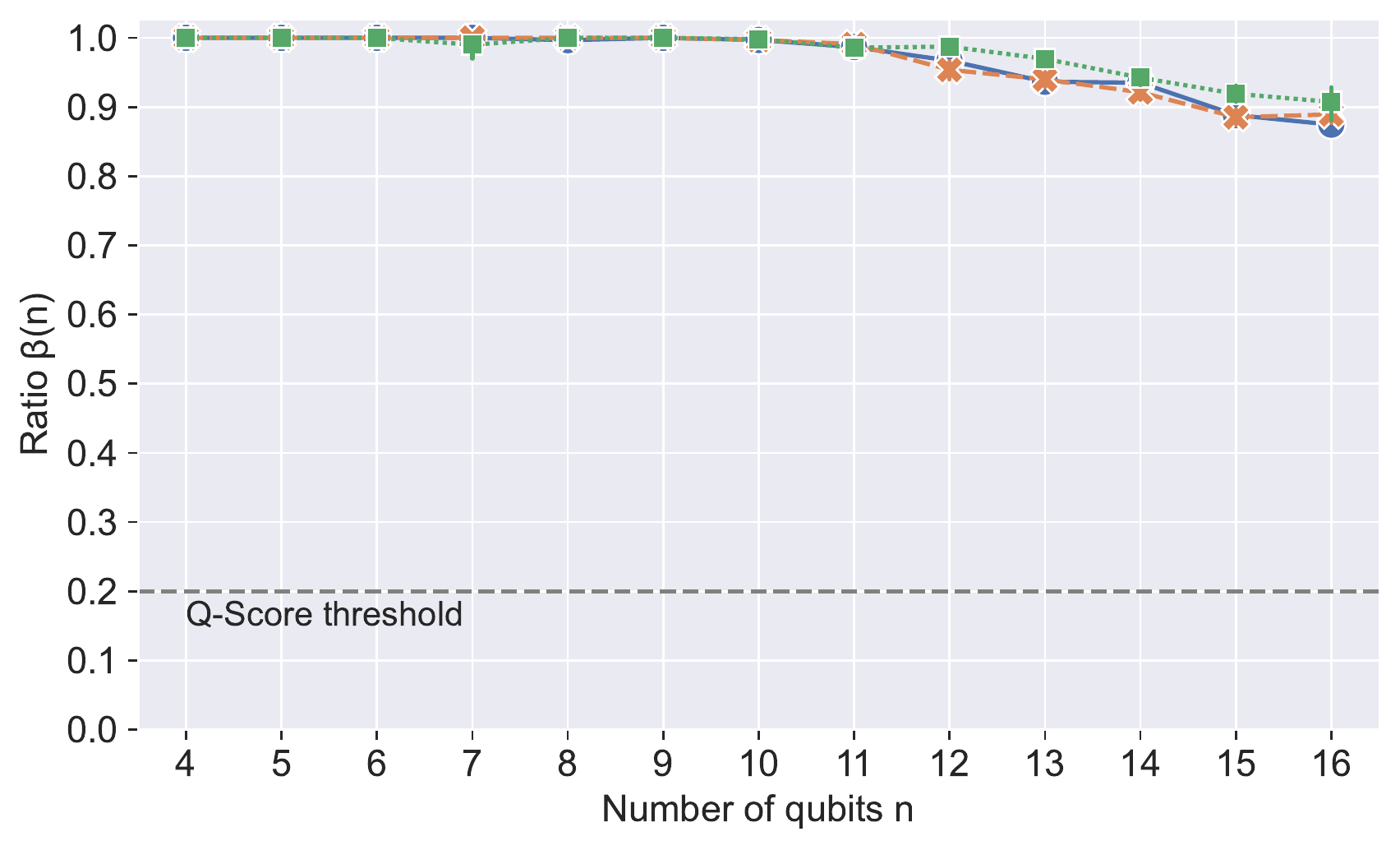} 
    \caption{Evolution of $\beta$(n) for MaxCut on non-UD graphs.}
    \label{ML_MC-non-ud}
  \end{subfigure}
  \begin{subfigure}[b]{0.49\linewidth}
    \centering 
    \captionsetup{width=1\linewidth}
    \includegraphics[width=\linewidth]{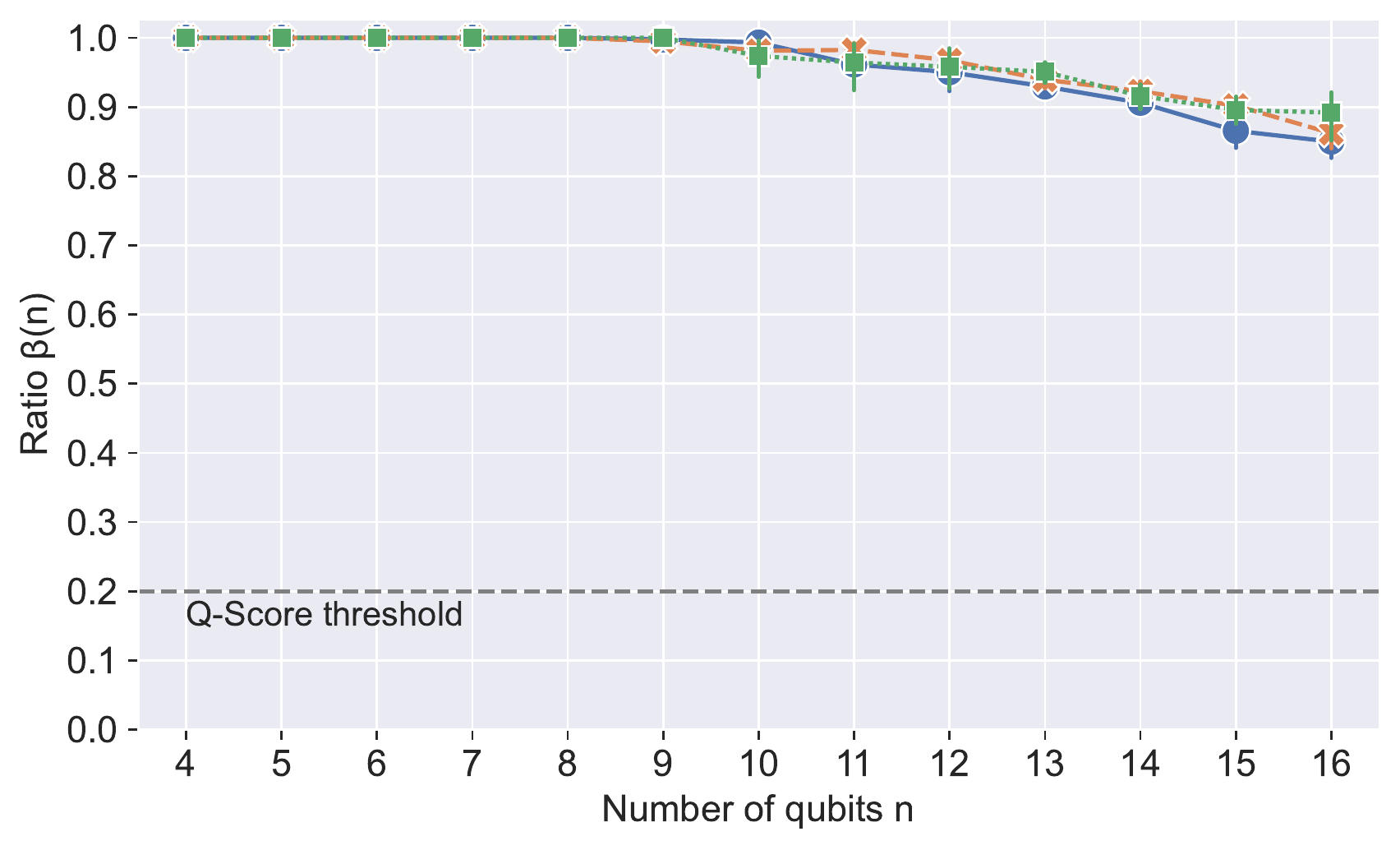} 
        \caption{Evolution of $\beta$(n) for MaxCut on UD graphs.}
    \label{ML_MC-ud}
  \end{subfigure}\captionsetup{justification=Justified}
      \begin{subfigure}[b]{0.49\linewidth}
    \centering
    \captionsetup{width=1\linewidth}
    \includegraphics[width=\linewidth]{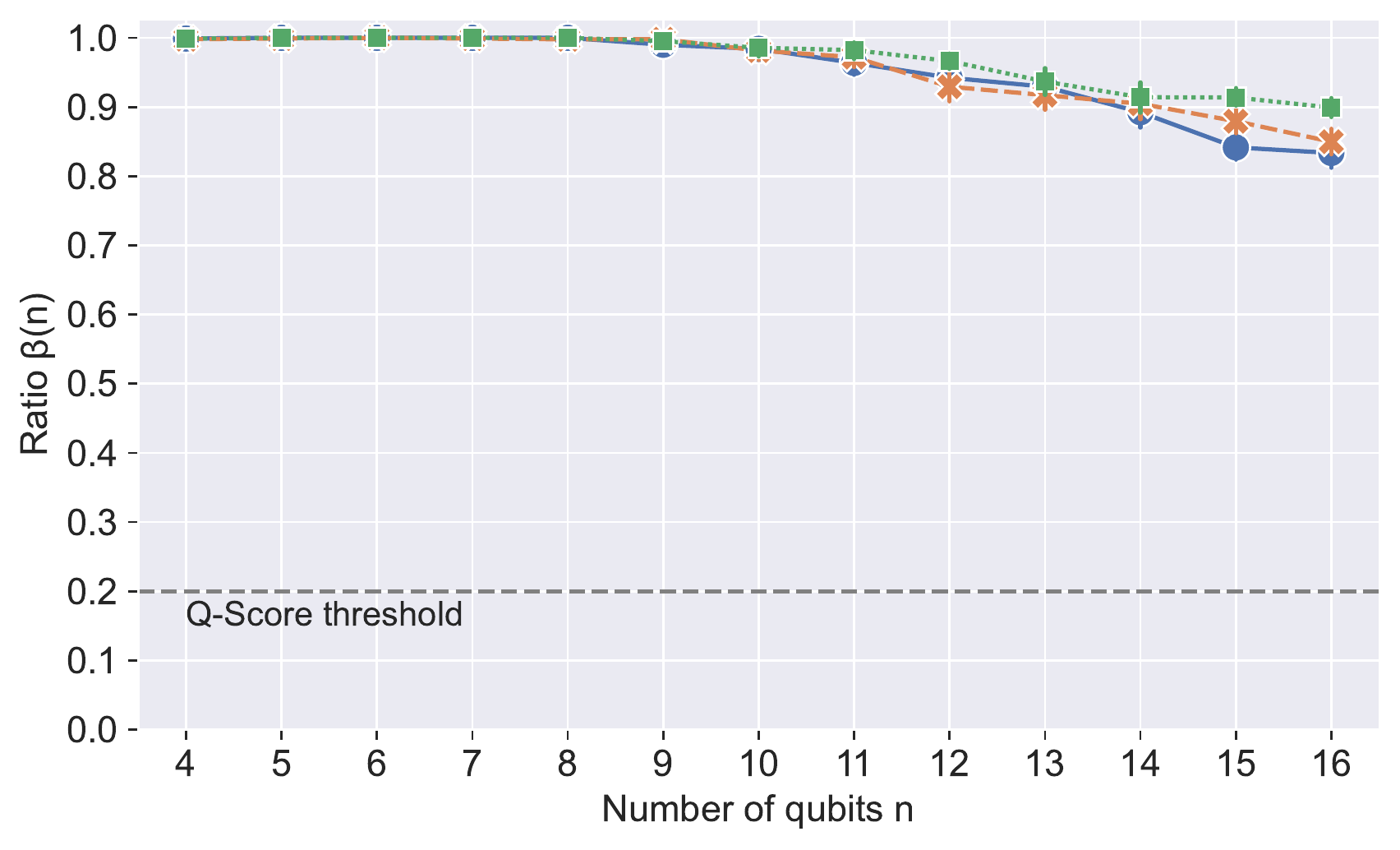} 
    \caption{Evolution of $\beta$(n) for MIS on non-UD graphs.}
    \label{ML_MIS-non-ud}
  \end{subfigure}
  \begin{subfigure}[b]{0.49\linewidth}
    \centering 
    \captionsetup{width=1\linewidth}
    \includegraphics[width=\linewidth]{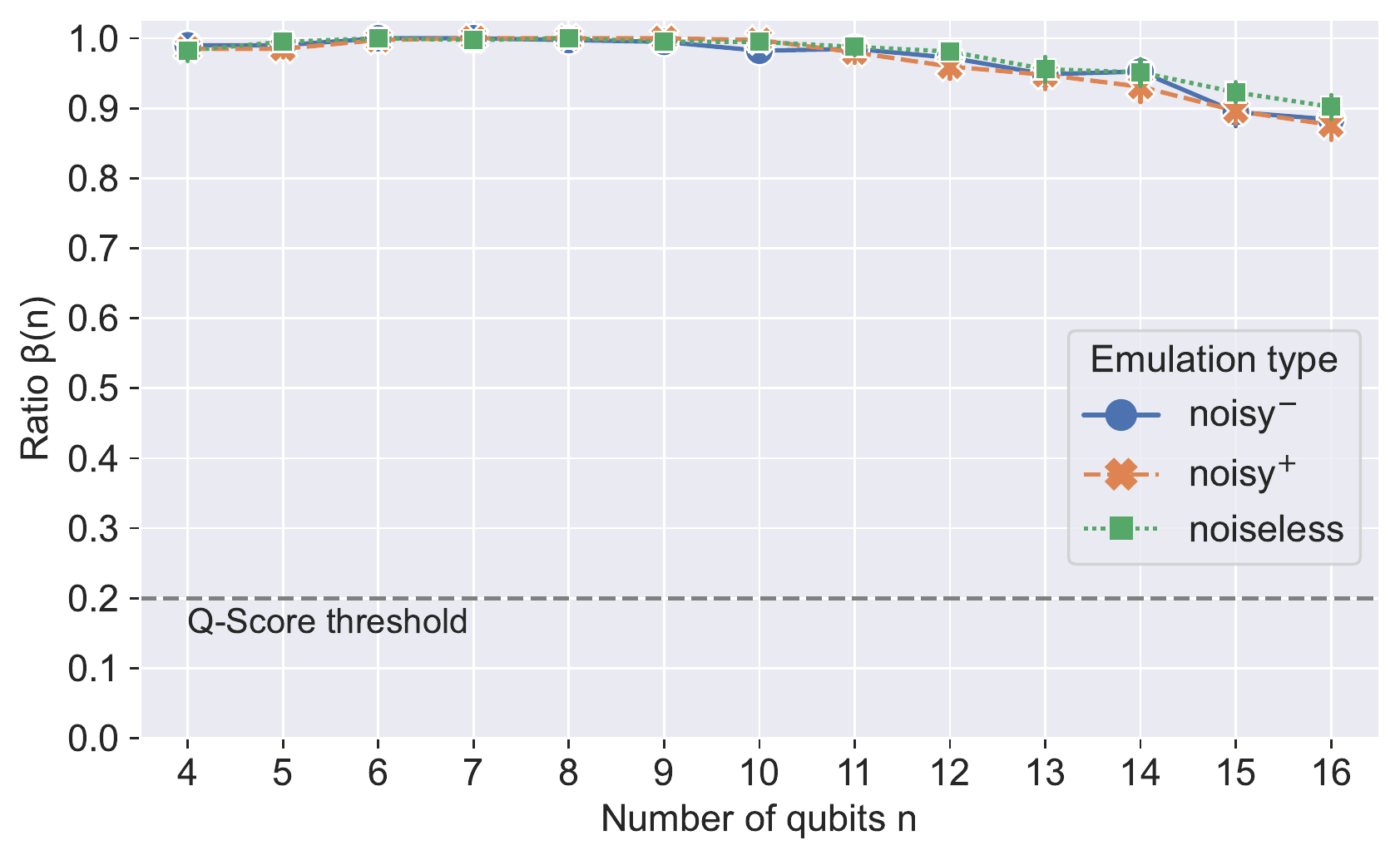} 
        \caption{Evolution of $\beta$(n) for MIS on UD graphs.}
    \label{ML_MIS-ud}
  \end{subfigure}\captionsetup{justification=Justified}
  \caption{Trained machine learning model: evolution of $\beta$(n) on different graph classes for the MaxCut and MIS problems.}
      \label{ML_beta}
\end{figure*}

Fig.~\ref{ML_beta} depicts the evolution of $\beta(n)$ (average and standard deviation with a 95\% confidence interval) on different graph classes ({\sl i.e.}, UD and non-UD graphs) for noisy and noiseless emulations. We first observe that the trained CMTRA could predict with high quality the solution for each instance. For instance, our proposed approach could always find the best solution for small and medium graphs (less than 12 nodes), even on noisy emulation (see Fig.~\ref{ML_MC-non-ud}). Also, the $\beta$ ratio was always above 0.85, even with noisy setups. For instance, the average $\beta$ on non-UD (resp. unit-disk) graphs with 16 nodes is roughly 0.89 when the \textit{noisy}$^+$ (resp. noiseless) model was applied. Moreover, we only observed a small impact of the noise on the bigger instances. Compared to the \textit{noiseless} emulations, the achieved $\beta$ on 16-node UD (resp. 12-node non-UD) graphs was reduced by approximately 5\% (resp. 2\%) when the \textit{noisy}$^-$ (resp. \textit{noisy}$^+$) emulations were run (see Figures~\ref{ML_MC-ud} and~\ref{ML_MC-non-ud}, respectively). This behavior might partially be explained by the fact that the CMTRA was trained with non-UD instances solved in a noiseless environment. Indeed, including UD-graphs with their related near-optimal pulses in noisy environments in the TDS might potentially improve the overall performance of the proposed CMTRA. Finally, we did not observe any important quality deterioration of the solutions found by the machine learning-based approach related to the class of the instance. For instance, the average $\beta$ ratio on non-UD graphs was always within the standard deviation range of unit-disk graphs' ratio for any register size. Also, no significant difference is observed when \textit{noisy}$^+$ setup is compared to \textit{noisy}$^-$. 

\subsubsection{MIS problem}
Figures~\ref{ML_MIS-non-ud} and \ref{ML_MIS-ud} show the evolution of $\beta$(n) on different graph classes for noisy and noiseless emulations for the MIS problem. First, we observed that the proposed CMTRA had a better performance on UD graphs (see Fig.~\ref{ML_MIS-ud}), especially when noisy models were applied. This behavior is expected because, as discussed in Section~\ref{methods}, unit-disk graphs are naturally embedded into neutral atom-based QPU, and the ground state of the related Hamiltonian can encode the solution to MIS on this class of graphs. Moreover, while noisy models did not impact the quality of the solutions on small and medium non-UD graphs (see Fig.~\ref{ML_MIS-non-ud}), the related $\beta$ on graphs with 15 (resp. 16) nodes was reduced by 6\% (resp. 5\%) when \textit{noisy}$^-$ (resp. \textit{noisy}$^+$) setup was applied.  

\subsection{Q-score estimation}

\begin{figure}[t] 
      \begin{subfigure}[b]{1\linewidth}
    \centering
    \captionsetup{width=1\linewidth}
    \includegraphics[width=\linewidth]{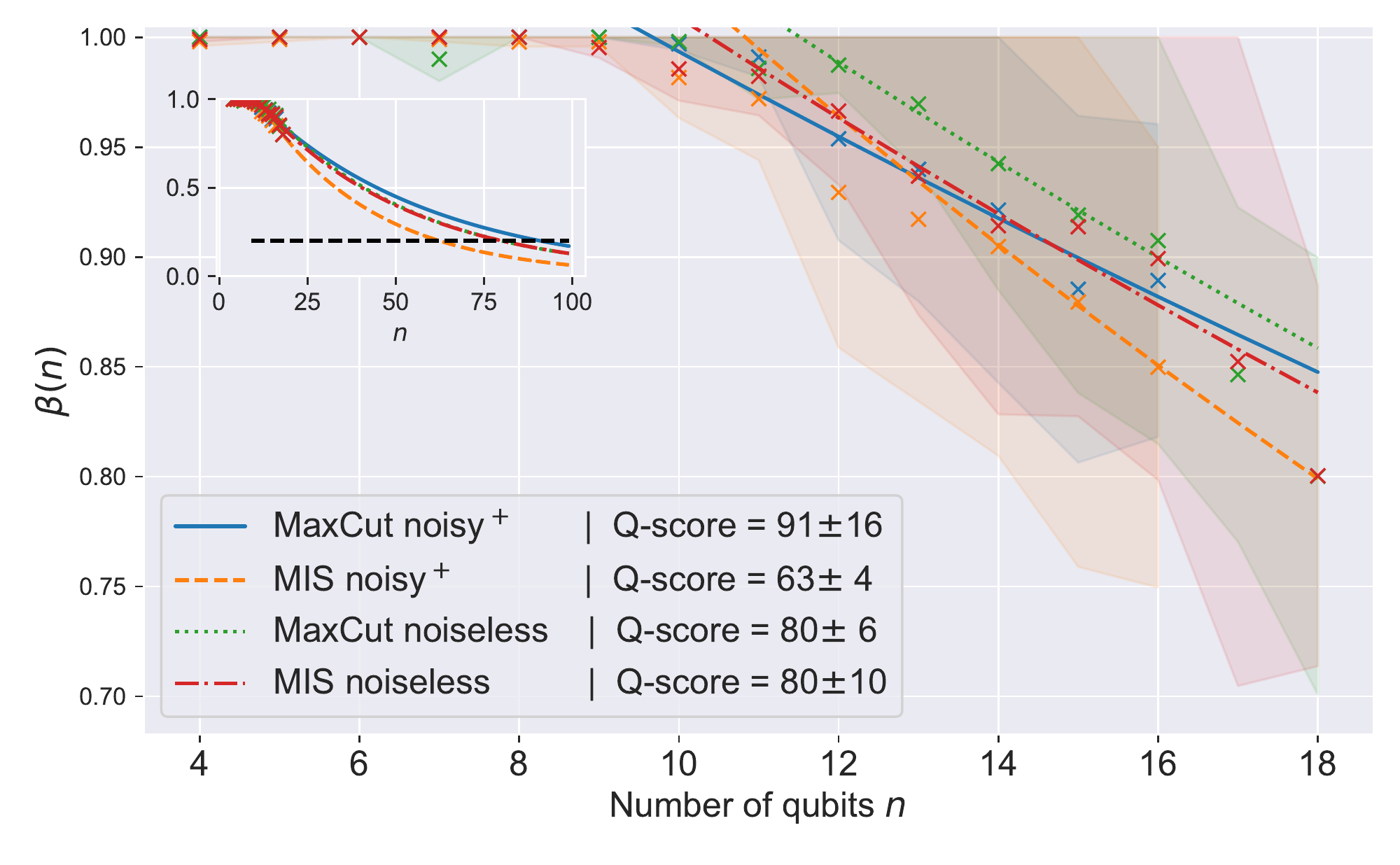} 
    \caption{Non-UD graphs.}
    \label{fig:extrapolation_nonUD}
  \end{subfigure}
  \begin{subfigure}[b]{1\linewidth}
    \centering 
    \captionsetup{width=1\linewidth}
    \includegraphics[width=\linewidth]{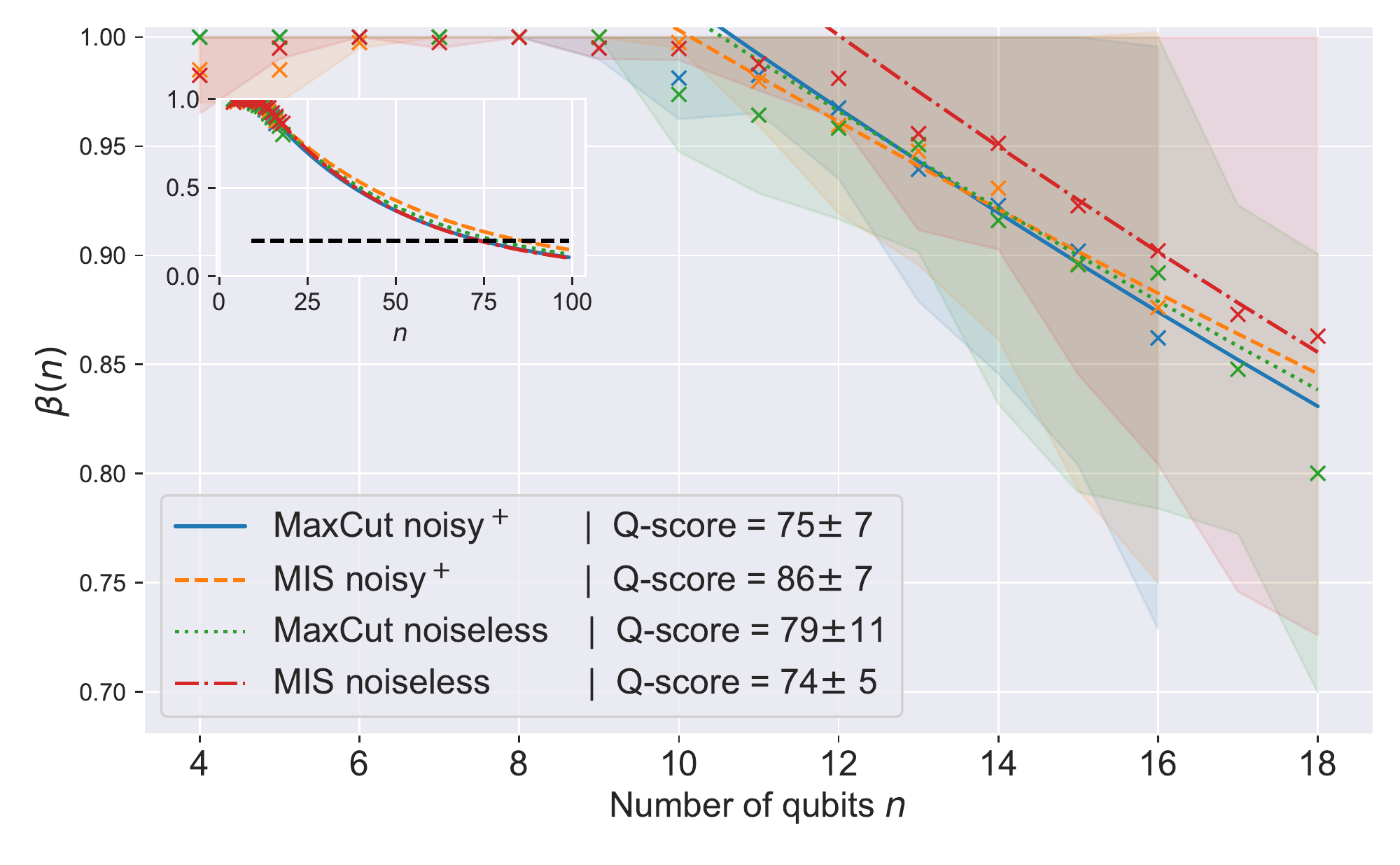} 
        \caption{UD graphs.}
    \label{fig:extrapolation_UD}
  \end{subfigure}\captionsetup{justification=Justified}
  \caption{Evaluation of the Q-score through the fit of an exponential decay of the scores. The value obtained is roughly the same in all cases, of the order of 80.}
      \label{fig:extrapolation}
\end{figure}

\begin{table}[b]\footnotesize
\begin{tabular}{cc|cc|}
\cline{3-4}
\multicolumn{2}{c|}{}                                   & \multicolumn{1}{c|}{~Noiseless~} &~ Noisy$^+$~ \\ \hline
\multicolumn{1}{|c|}{\multirow{2}{*}{MIS}}     & UD graphs     & \multicolumn{1}{c|}{$74\pm5$}         & $86\pm7$    \\ \cline{2-4} 
\multicolumn{1}{|c|}{}                         & ~ Non-UD graphs~ & \multicolumn{1}{c|}{$80\pm10$}         & $63\pm4$    \\ \hline
\multicolumn{1}{|c|}{\multirow{2}{*}{MaxCut}}  & UD graphs     & \multicolumn{1}{c|}{$79\pm11$}        & $75\pm7$    \\ \cline{2-4} 
\multicolumn{1}{|c|}{}                         & Non-UD graphs& \multicolumn{1}{c|}{$80\pm6$}         & $91\pm16$    \\ \hline
\end{tabular}\captionsetup{justification=Justified}
\caption{Estimated Q-scores for MIS and MaxCut problems on Unit-Disk and non-UD graphs and in a noisy$^+$ and noiseless settings.}
\label{tab:summary}
\end{table}

As discussed in Section \ref{methods}, the Q-score is defined as the problem size at which the score of the proposed quantum algorithm becomes less than 20\% better than a random sampling. 
In its original definition, the average score expected from the final state of the qubits was used. Even though this choice captures the overall {\sl quality}
of the final state, it does not give any information about the distribution of scores. In particular, it does not say how the score is expected to improve as the
number of samples of the state is increased.
Because our method directly provides a pulse shape, all runs on the QPU will be preparing the same state (up to noise). We can then afford to increase the number
of samples.
We then defined our score in a way that is closer to how it would be used in practice: given a number of allowed samples (or {\sl time budget}), what is the best
sampled solution? Here, we chose 1000 samples for each graph\footnote{Note that for small graph sizes, this would allow an almost extensive sampling of the Hilbert space.}, 
independently of its size. Since the computing time on the analog QPU is independent of the number of qubits, this is equivalent to setting a fixed total computation time.
This number was found to be a good compromise between reasonably short computation time and good enough performances.

In our results, the score obtained stayed above the 20\%, even in the presence of noise, up to the largest graphs we were able to simulate with noise in a reasonable amount of time.
In order to determine the Q-score of the method and platform we need to extrapolate the results to larger problem sizes.
To this end, we fit an exponential decay on the tail of the size dependence of the score $\beta(n) = \beta_0 e^{-n/n_0}$. The Q-score is then given by $Q_{score}=n_0\log(5\beta_0)$. The results are shown in the inset of Fig.~\ref{fig:extrapolation} and summarized in \ref{tab:summary}. For both problems, the Q-score is of the order of 80 (except for MIS of non-UD graphs on noisy devices), to be compared with the Q-score determined in~\cite{qscore} for QAOA on state-of-the-art gate-based QC platforms. 
In particular, the presence of noise does not seem to lower significantly the score. Indeed, this specific behavior of analog quantum computing is very different from what was observed in the digital quantum circuits~\cite{qscore}, where the score degrades faster for larger circuit depths.
The comparison between the two approaches is not easy, as there is no equivalent to the circuit depth here.
However, this example highlights the resilience to noises of the analog approach.
The score is then expected to be mainly coming from the quality of the training data set as well as the difficulty to define a proper embedding of the problem. The former can be improved upon by additional optimization steps in the training set, as well as by including data from larger graphs (coming either from lengthier emulation or actual QPU results). Concerning the latter, significant improvement can be expected from well-thought embedding heuristics, as illustrated in Fig.~\ref{fig:easier_graph}, but this would demand a deeper analysis of the data set, and it might end up being an intrinsic limitation of the method.

One may observe that, even though the Q-score on the noisy device was sometimes found to be larger than the corresponding noiseless approach, its true value should be smaller. Indeed, as the estimated error on the value suggests, this estimation was mostly qualitative. If one were to push the analysis and the training further, both the noiseless and the noisy would get a higher Q-score and one would expect the noiseless to perform better. It is worth noting, however, that there are cases where noise (and non-unitary dynamics) can provide a benefit~\cite{Novo_2018}, as the thermal fluctuations can induce a deconfinement of excitations~\cite{Henry_2014} and therefore help in the exploration of the Hilbert space.
\section*{Conclusion} 
In this study, we demonstrated that it is possible, thanks to machine learning, to develop an efficient way of solving combinatorial graph problems on analog quantum processing units such as neutral atom platforms.
Determining a {\sl good} pulse, as well as a {\sl good} embedding, to solve the problem on a given graph is one of the most important bottlenecks of the analog approach. By providing directly a good pulse, our method allows restricting the runs on the Quantum device to the sampling of a given final state, reducing dramatically the number of shots and hence the time-to-solution.
We showed that it is possible to train a model to predict a pulse that prepares a final state with a sufficient overlap of (near-) optimal solutions of the problem's instances.

In this study, we chose a training set that was not fully optimized, so that its generation would not take too long. If one would be to improve the performance, one step would be to improve the training set, both by pushing further optimization for each of its instances, as well as increasing the largest graph order ({\sl i.e.}, number of nodes) it contains. Also, the quality of the result depends on the number of shots of the final state one allows taking (the {\sl time budget}). One could also try to specifically train the model for a fixed number of shots. Alternatively, one could try a reinforcement learning scheme. In that case, the training is expected to take longer, but one would spare the generation of the training data set. A similar approach could be applied to other combinatorial graph problems.

Furthermore, each model we trained in this study had its own embedding strategy. As highlighted in Fig.~\ref{fig:easier_graph}, smarter embedding schemes can yield significantly better results. It is worth mentioning that, because of the difficulty of embedding a generic graph, it may be more efficient to use an alternative representation of the initial problem. For example, instead of solving the MIS problem on a graph, one could solve the equivalent Maximum Clique problem on the complement graph. 
  
Our results highlight the potential of NISQ-era, analog quantum computing. Even though the need to develop problem-specific frameworks may seem to contradict the goal of speeding-up classical calculations, it could lead to the first quantum advantageous solution.

\section*{Acknowledgments}

We thank Thomas Ayral, Lucas Leclerc, Vincent Elfving, and Loïc Henriet for insightful discussions.

\bibliographystyle{bib_sobraep}
\bibliography{refers}

\begin{thebibliography}{10}
\newcommand{\enquote}[1]{``#1''}
\providecommand{\url}[1]{{\tt #1}}
\providecommand{\urlprefix}{URL: }
\expandafter\ifx\csname urlstyle\endcsname\relax
  \providecommand{\doi}[1]{doi:\discretionary{}{}{}#1}\else
  \providecommand{\doi}{doi:\discretionary{}{}{}\begingroup
  \urlstyle{rm}\Url}\fi
\providecommand{\eprint}[2][]{\url{#2}}

\bibitem{PRXQuantum.2.020343}
I.~Pogorelov, T.~Feldker, C.~D. Marciniak, L.~Postler, G.~Jacob,
  O.~Krieglsteiner, V.~Podlesnic, M.~Meth, V.~Negnevitsky, M.~Stadler,
  B.~H\"ofer, C.~W\"achter, K.~Lakhmanskiy, R.~Blatt, P.~Schindler, T.~Monz,
  \enquote{Compact Ion-Trap Quantum Computing Demonstrator}, {\em PRX
  Quantum\/}, vol.~2, p. 020343, Jun 2021, \doi{10.1103/PRXQuantum.2.020343},
  \urlprefix\url{https://link.aps.org/doi/10.1103/PRXQuantum.2.020343}.

\bibitem{Benhelm_2008}
J.~Benhelm, G.~Kirchmair, C.~F. Roos, R.~Blatt, \enquote{Towards fault-tolerant
  quantum computing with trapped ions}, {\em Nature Physics\/}, vol.~4, no.~6,
  pp. 463--466, apr 2008, \doi{10.1038/nphys961},
  \urlprefix\url{https://doi.org/10.1038%2Fnphys961}.

\bibitem{Kjaergaard2020}
M.~Kjaergaard, M.~E. Schwartz, J.~Braum\"{u}ller, P.~Krantz, J.~I.-J. Wang,
  S.~Gustavsson, W.~D. Oliver, \enquote{Superconducting Qubits: Current State
  of Play}, {\em Annual Review of Condensed Matter Physics\/}, vol.~11, no.~1,
  pp. 369--395, 2020, \doi{10.1146/annurev-conmatphys-031119-050605},
  \urlprefix\url{https://doi.org/10.1146/annurev-conmatphys-031119-050605},
  \eprint{https://doi.org/10.1146/annurev-conmatphys-031119-050605}.

\bibitem{Devoret2004}
M.~H. Devoret, A.~Wallraff, J.~M. Martinis, \enquote{Superconducting Qubits: A
  Short Review}, , 2004, \doi{10.48550/ARXIV.COND-MAT/0411174},
  \urlprefix\url{https://arxiv.org/abs/cond-mat/0411174}.

\bibitem{henriet2020quantum}
L.~Henriet, L.~Beguin, A.~Signoles, T.~Lahaye, A.~Browaeys, G.-O. Reymond,
  C.~Jurczak, \enquote{Quantum computing with neutral atoms}, {\em Quantum\/},
  vol.~4, p. 327, 2020.

\bibitem{wurtz2022}
J.~Wurtz, P.~Lopes, N.~Gemelke, A.~Keesling, S.~Wang, \enquote{Industry
  applications of neutral-atom quantum computing solving independent set
  problems}, , 2022, \doi{10.48550/ARXIV.2205.08500},
  \urlprefix\url{https://arxiv.org/abs/2205.08500}.

\bibitem{Qgates}
C.~P. Williams, {\em Quantum Gates\/}, pp. 51--122, Springer London, London,
  2011, \doi{10.1007/978-1-84628-887-6_2},
  \urlprefix\url{https://doi.org/10.1007/978-1-84628-887-6_2}.

\bibitem{Preskill_2018}
J.~Preskill, \enquote{Quantum Computing in the {NISQ} era and beyond}, {\em
  Quantum\/}, vol.~2, p.~79, aug 2018, \doi{10.22331/q-2018-08-06-79},
  \urlprefix\url{https://doi.org/10.22331%2Fq-2018-08-06-79}.

\bibitem{PhysRevA.100.032328}
A.~W. Cross, L.~S. Bishop, S.~Sheldon, P.~D. Nation, J.~M. Gambetta,
  \enquote{Validating quantum computers using randomized model circuits}, {\em
  Phys Rev A\/}, vol. 100, p. 032328, Sep 2019,
  \doi{10.1103/PhysRevA.100.032328},
  \urlprefix\url{https://link.aps.org/doi/10.1103/PhysRevA.100.032328}.

\bibitem{qscore}
S.~Martiel, T.~Ayral, C.~Allouche, \enquote{Benchmarking Quantum Coprocessors
  in an Application-Centric, Hardware-Agnostic, and Scalable Way}, {\em IEEE
  Transactions on Quantum Engineering\/}, vol.~2, pp. 1--11, 2021,
  \doi{10.1109/TQE.2021.3090207}.

\bibitem{Bharti_2022}
K.~Bharti, A.~Cervera-Lierta, T.~H. Kyaw, T.~Haug, S.~Alperin-Lea, A.~Anand,
  M.~Degroote, H.~Heimonen, J.~S. Kottmann, T.~Menke, W.-K. Mok, S.~Sim, L.-C.
  Kwek, A.~Aspuru-Guzik, \enquote{Noisy intermediate-scale quantum algorithms},
  {\em Reviews of Modern Physics\/}, vol.~94, no.~1, feb 2022,
  \doi{10.1103/revmodphys.94.015004},
  \urlprefix\url{https://doi.org/10.1103%2Frevmodphys.94.015004}.

\bibitem{Nishi_2021}
H.~Nishi, T.~Kosugi, Y.~ichiro Matsushita, \enquote{Implementation of quantum
  imaginary-time evolution method on {NISQ} devices by introducing nonlocal
  approximation}, {\em npj Quantum Information\/}, vol.~7, no.~1, jun 2021,
  \doi{10.1038/s41534-021-00409-y},
  \urlprefix\url{https://doi.org/10.1038%2Fs41534-021-00409-y}.

\bibitem{Novo_2018}
L.~Novo, S.~Chakraborty, M.~Mohseni, Y.~Omar, \enquote{Environment-assisted
  analog quantum search}, {\em Physical Review A\/}, vol.~98, no.~2, aug 2018,
  \doi{10.1103/physreva.98.022316},
  \urlprefix\url{https://doi.org/10.1103%2Fphysreva.98.022316}.

\bibitem{graph_rydberg1}
J.~Wurtz, P.~Lopes, N.~Gemelke, A.~Keesling, S.~Wang, \enquote{Industry
  applications of neutral-atom quantum computing solving independent set
  problems}, , 2022, \doi{10.48550/ARXIV.2205.08500},
  \urlprefix\url{https://arxiv.org/abs/2205.08500}.

\bibitem{graph_rydberg2}
S.~Ebadi, A.~Keesling, M.~Cain, T.~T. Wang, H.~Levine, D.~Bluvstein,
  G.~Semeghini, A.~Omran, J.-G. Liu, R.~Samajdar, X.-Z. Luo, B.~Nash, X.~Gao,
  B.~Barak, E.~Farhi, S.~Sachdev, N.~Gemelke, L.~Zhou, S.~Choi, H.~Pichler,
  S.-T. Wang, M.~Greiner, V.~Vuleti{\'{c} }, M.~D. Lukin, \enquote{Quantum
  optimization of maximum independent set using Rydberg atom arrays}, {\em
  Science\/}, vol. 376, no. 6598, pp. 1209--1215, jun 2022,
  \doi{10.1126/science.abo6587},
  \urlprefix\url{https://doi.org/10.1126%2Fscience.abo6587}.

\bibitem{graph_rydberg3}
M.~Kim, K.~Kim, J.~Hwang, E.-G. Moon, J.~Ahn, \enquote{Rydberg Quantum Wires
  for Maximum Independent Set Problems with Nonplanar and High-Degree Graphs},
  , 2021, \doi{10.48550/ARXIV.2109.03517},
  \urlprefix\url{https://arxiv.org/abs/2109.03517}.

\bibitem{dalyac2021qualifying}
C.~Dalyac, L.~Henriet, E.~Jeandel, W.~Lechner, S.~Perdrix, M.~Porcheron,
  M.~Veshchezerova, \enquote{Qualifying quantum approaches for hard industrial
  optimization problems. A case study in the field of smart-charging of
  electric vehicles}, {\em EPJ Quantum Technology\/}, vol.~8, no.~1, p.~12,
  2021.

\bibitem{QA1}
A.~Rajak, S.~Suzuki, A.~Dutta, B.~K. Chakrabarti, \enquote{Quantum Annealing:
  An Overview}, , 2022, \doi{10.48550/ARXIV.2207.01827},
  \urlprefix\url{https://arxiv.org/abs/2207.01827}.

\bibitem{bondy1976graph}
J.~A. Bondy, U.~S.~R. Murty, {\em et~al.\/}, {\em Graph theory with
  applications\/}, vol. 290, Macmillan London, 1976.

\bibitem{9625601}
A.~Kershenbaum, {\em Telecommunications network design algorithms\/},
  McGraw-Hill, Inc., 1993.

\bibitem{roth2010suggesting}
M.~Roth, A.~Ben-David, D.~Deutscher, G.~Flysher, I.~Horn, A.~Leichtberg,
  N.~Leiser, Y.~Matias, R.~Merom, \enquote{Suggesting friends using the
  implicit social graph}, {\em in\/} {\em Proceedings of the 16th ACM SIGKDD
  international conference on Knowledge discovery and data mining\/}, pp.
  233--242, 2010.

\bibitem{barahona1988application}
F.~Barahona, M.~Gr{\"o}tschel, M.~J{\"u}nger, G.~Reinelt, \enquote{An
  application of combinatorial optimization to statistical physics and circuit
  layout design}, {\em Operations Research\/}, vol.~36, no.~3, pp. 493--513,
  1988.

\bibitem{karp1972reducibility}
R.~M. Karp, \enquote{Reducibility among combinatorial problems}, {\em in\/}
  {\em Complexity of computer computations\/}, pp. 85--103, Springer, 1972.

\bibitem{berman1994approximating}
P.~Berman, M.~F{\"u}rer, \enquote{Approximating Maximum Independent Set}, {\em
  in\/} {\em Proceedings of the fifth annual ACM-SIAM Symposium on discrete
  algorithms\/}, 70, p. 365, SIAM, 1994.

\bibitem{goemans1994879}
M.~X. Goemans, D.~P. Williamson, \enquote{. 879-approximation algorithms for
  max cut and max 2sat}, {\em in\/} {\em Proceedings of the twenty-sixth annual
  ACM symposium on Theory of computing\/}, pp. 422--431, 1994.

\bibitem{xiao2017exact}
M.~Xiao, H.~Nagamochi, \enquote{Exact algorithms for maximum independent set},
  {\em Information and Computation\/}, vol. 255, pp. 126--146, 2017.

\bibitem{rendl2010solving}
F.~Rendl, G.~Rinaldi, A.~Wiegele, \enquote{Solving max-cut to optimality by
  intersecting semidefinite and polyhedral relaxations}, {\em Mathematical
  Programming\/}, vol. 121, no.~2, pp. 307--335, 2010.

\bibitem{das2012heuristics}
K.~N. Das, B.~Chaudhuri, \enquote{Heuristics to find maximum independent set:
  An overview}, {\em in\/} {\em Proceedings of the International Conference on
  Soft Computing for Problem Solving (SocProS 2011) December 20-22, 2011\/},
  pp. 881--892, Springer, 2012.

\bibitem{kim2019comparison}
Y.-H. Kim, Y.~Yoon, Z.~W. Geem, \enquote{A comparison study of harmony search
  and genetic algorithm for the max-cut problem}, {\em Swarm and evolutionary
  computation\/}, vol.~44, pp. 130--135, 2019.

\bibitem{farhi2014quantum}
E.~Farhi, J.~Goldstone, S.~Gutmann, \enquote{A quantum approximate optimization
  algorithm}, {\em arXiv preprint arXiv:14114028\/}, 2014.

\bibitem{brandao2018fixed}
F.~G. Brandao, M.~Broughton, E.~Farhi, S.~Gutmann, H.~Neven, \enquote{For fixed
  control parameters the quantum approximate optimization algorithm's objective
  function value concentrates for typical instances}, {\em arXiv preprint
  arXiv:181204170\/}, 2018.

\bibitem{zhou2020quantum}
L.~Zhou, S.-T. Wang, S.~Choi, H.~Pichler, M.~D. Lukin, \enquote{Quantum
  approximate optimization algorithm: Performance, mechanism, and
  implementation on near-term devices}, {\em Physical Review X\/}, vol.~10,
  no.~2, p. 021067, 2020.

\bibitem{crooks2018performance}
G.~E. Crooks, \enquote{Performance of the quantum approximate optimization
  algorithm on the maximum cut problem}, {\em arXiv preprint
  arXiv:181108419\/}, 2018.

\bibitem{medvidovic2021classical}
M.~Medvidovi{\'c}, G.~Carleo, \enquote{Classical variational simulation of the
  quantum approximate optimization algorithm}, {\em npj Quantum Information\/},
  vol.~7, no.~1, pp. 1--7, 2021.

\bibitem{herrman2021impact}
R.~Herrman, L.~Treffert, J.~Ostrowski, P.~C. Lotshaw, T.~S. Humble, G.~Siopsis,
  \enquote{Impact of graph structures for QAOA on MaxCut}, {\em Quantum
  Information Processing\/}, vol.~20, no.~9, pp. 1--21, 2021.

\bibitem{Markovi__2020}
D.~Markovi{\'{c} }, J.~Grollier, \enquote{Quantum neuromorphic computing}, {\em
  Applied Physics Letters\/}, vol. 117, no.~15, p. 150501, oct 2020,
  \doi{10.1063/5.0020014}, \urlprefix\url{https://doi.org/10.1063%2F5.0020014}.

\bibitem{ZAK19991583}
M.~Zak, \enquote{Quantum Analog Computing}, {\em Chaos, Solitons \&
  Fractals\/}, vol.~10, no.~10, pp. 1583--1620, 1999,
  \doi{https://doi.org/10.1016/S0960-0779(98)00215-X},
  \urlprefix\url{https://www.sciencedirect.com/science/article/pii/S096007799800215X}.

\bibitem{Pichler}
H.~Pichler, S.-T. Wang, L.~Zhou, S.~Choi, M.~D. Lukin, \enquote{Quantum
  Optimization for Maximum Independent Set Using Rydberg Atom Arrays}, , 2018,
  \doi{10.48550/ARXIV.1808.10816},
  \urlprefix\url{https://arxiv.org/abs/1808.10816}.

\bibitem{ebadi2022quantum}
S.~Ebadi, A.~Keesling, M.~Cain, T.~T. Wang, H.~Levine, D.~Bluvstein,
  G.~Semeghini, A.~Omran, J.-G. Liu, R.~Samajdar, {\em et~al.\/},
  \enquote{Quantum optimization of maximum independent set using Rydberg atom
  arrays}, {\em Science\/}, p. eabo6587, 2022.

\bibitem{Barahona1982}
F.~Barahona, \enquote{On the computational complexity of Ising spin glass
  models}, {\em Journal of Physics A: Mathematical and General\/}, vol.~15,
  no.~10, pp. 3241--3253, oct 1982, \doi{10.1088/0305-4470/15/10/028},
  \urlprefix\url{https://doi.org/10.1088/0305-4470/15/10/028}.

\bibitem{yaacoby2022comparison}
R.~Yaacoby, N.~Schaar, L.~Kellerhals, O.~Raz, D.~Hermelin, R.~Pugatch,
  \enquote{Comparison between a quantum annealer and a classical approximation
  algorithm for computing the ground state of an Ising spin glass}, {\em
  Physical Review E\/}, vol. 105, no.~3, p. 035305, 2022.

\bibitem{coja2022ising}
A.~Coja-Oghlan, P.~Loick, B.~F. Mezei, G.~B. Sorkin, \enquote{The Ising
  antiferromagnet and max cut on random regular graphs}, {\em SIAM Journal on
  Discrete Mathematics\/}, vol.~36, no.~2, pp. 1306--1342, 2022.

\bibitem{zhang2020computational}
Z.~Zhang, \enquote{Computational complexity of spin-glass three-dimensional
  (3D) Ising model}, {\em Journal of Materials Science \& Technology\/},
  vol.~44, pp. 116--120, 2020.

\bibitem{Fu1986}
Y.~Fu, P.~W. Anderson, \enquote{Application of statistical mechanics to
  {NP}-complete problems in combinatorial optimisation}, {\em Journal of
  Physics A: Mathematical and General\/}, vol.~19, no.~9, pp. 1605--1620, jun
  1986, \doi{10.1088/0305-4470/19/9/033},
  \urlprefix\url{https://doi.org/10.1088/0305-4470/19/9/033}.

\bibitem{Mezard1986}
M.~Mezard, G.~Parisi, M.~Virasoro, {\em Spin Glass Theory and Beyond\/}, WORLD
  SCIENTIFIC, 1986, \doi{10.1142/0271},
  \urlprefix\url{https://www.worldscientific.com/doi/abs/10.1142/0271},
  \eprint{https://www.worldscientific.com/doi/pdf/10.1142/0271}.

\bibitem{Hartmann2005}
A.~Hartmann, M.~Weigt, {\em Phase Transitions in Combinatorial Optimization
  Problems: Basics, Algorithms and Statistical Mechanics\/}, John Wiley \&
  Sons, Ltd, 12005, \doi{10.1002/3527606734},
  \urlprefix\url{https://onlinelibrary.wiley.com/doi/book/10.1002/3527606734}.

\bibitem{Henry_2021}
L.-P. Henry, S.~Thabet, C.~Dalyac, L.~Henriet, \enquote{Quantum evolution
  kernel: Machine learning on graphs with programmable arrays of qubits}, {\em
  Physical Review A\/}, vol. 104, no.~3, sep 2021,
  \doi{10.1103/physreva.104.032416},
  \urlprefix\url{https://doi.org/10.1103%2Fphysreva.104.032416}.

\bibitem{Farhi2001}
E.~Farhi, J.~Goldstone, S.~Gutmann, J.~Lapan, A.~Lundgren, D.~Preda, \enquote{A
  Quantum Adiabatic Evolution Algorithm Applied to Random Instances of an
  NP-Complete Problem}, {\em Science\/}, vol. 292, no. 5516, pp. 472--475,
  2001, \doi{10.1126/science.1057726},
  \urlprefix\url{https://www.science.org/doi/abs/10.1126/science.1057726},
  \eprint{https://www.science.org/doi/pdf/10.1126/science.1057726}.

\bibitem{exponentialgap}
N.~G. Dickson, M.~H.~S. Amin, \enquote{Does Adiabatic Quantum Optimization Fail
  for {NP}-Complete Problems?}, {\em Physical Review Letters\/}, vol. 106,
  no.~5, feb 2011, \doi{10.1103/physrevlett.106.050502},
  \urlprefix\url{https://doi.org/10.1103%2Fphysrevlett.106.050502}.

\bibitem{noisemodel}
S.~de~L\'es\'eleuc, D.~Barredo, V.~Lienhard, A.~Browaeys, T.~Lahaye,
  \enquote{Analysis of imperfections in the coherent optical excitation of
  single atoms to Rydberg states}, {\em Phys Rev A\/}, vol.~97, p. 053803, May
  2018, \doi{10.1103/PhysRevA.97.053803},
  \urlprefix\url{https://link.aps.org/doi/10.1103/PhysRevA.97.053803}.

\bibitem{breu1998unit}
H.~Breu, D.~G. Kirkpatrick, \enquote{Unit disk graph recognition is NP-hard},
  {\em Computational Geometry\/}, vol.~9, no. 1-2, pp. 3--24, 1998.

\bibitem{fruchterman1991graph}
T.~M. Fruchterman, E.~M. Reingold, \enquote{Graph drawing by force-directed
  placement}, {\em Software: Practice and experience\/}, vol.~21, no.~11, pp.
  1129--1164, 1991.

\bibitem{Khairy_2020}
S.~Khairy, R.~Shaydulin, L.~Cincio, Y.~Alexeev, P.~Balaprakash,
  \enquote{Learning to Optimize Variational Quantum Circuits to Solve
  Combinatorial Problems}, {\em Proceedings of the {AAAI} Conference on
  Artificial Intelligence\/}, vol.~34, no.~03, pp. 2367--2375, apr 2020,
  \doi{10.1609/aaai.v34i03.5616},
  \urlprefix\url{https://doi.org/10.1609%2Faaai.v34i03.5616}.

\bibitem{khairy2019reinforcement}
S.~Khairy, R.~Shaydulin, L.~Cincio, Y.~Alexeev, P.~Balaprakash,
  \enquote{Reinforcement-learning-based variational quantum circuits
  optimization for combinatorial problems}, {\em arXiv preprint
  arXiv:191104574\/}, 2019.

\bibitem{mason1999boosting}
L.~Mason, J.~Baxter, P.~Bartlett, M.~Frean, \enquote{Boosting algorithms as
  gradient descent}, {\em Advances in neural information processing systems\/},
  vol.~12, 1999.

\bibitem{frazier2018bayesian}
P.~I. Frazier, \enquote{Bayesian optimization}, {\em in\/} {\em Recent advances
  in optimization and modeling of contemporary problems\/}, pp. 255--278,
  Informs, 2018.

\bibitem{demirel2019ensemble}
K.~C. Demirel, A.~Sahin, E.~Albey, \enquote{Ensemble Learning based on
  Regressor Chains: A Case on Quality Prediction.}, {\em in\/} {\em DATA\/},
  pp. 267--274, 2019.

\bibitem{friedman2001greedy}
J.~H. Friedman, \enquote{Greedy function approximation: a gradient boosting
  machine}, {\em Annals of statistics\/}, pp. 1189--1232, 2001.

\bibitem{RB2011}
E.~Magesan, J.~M. Gambetta, J.~Emerson, \enquote{Scalable and Robust Randomized
  Benchmarking of Quantum Processes}, {\em Phys Rev Lett\/}, vol. 106, p.
  180504, May 2011, \doi{10.1103/PhysRevLett.106.180504},
  \urlprefix\url{https://link.aps.org/doi/10.1103/PhysRevLett.106.180504}.

\bibitem{RB2012}
E.~Magesan, J.~M. Gambetta, J.~Emerson, \enquote{Characterizing quantum gates
  via randomized benchmarking}, {\em Phys Rev A\/}, vol.~85, p. 042311, Apr
  2012, \doi{10.1103/PhysRevA.85.042311},
  \urlprefix\url{https://link.aps.org/doi/10.1103/PhysRevA.85.042311}.

\bibitem{QPT2013}
S.~T. Merkel, J.~M. Gambetta, J.~A. Smolin, S.~Poletto, A.~D. C\'orcoles, B.~R.
  Johnson, C.~A. Ryan, M.~Steffen, \enquote{Self-consistent quantum process
  tomography}, {\em Phys Rev A\/}, vol.~87, p. 062119, Jun 2013,
  \doi{10.1103/PhysRevA.87.062119},
  \urlprefix\url{https://link.aps.org/doi/10.1103/PhysRevA.87.062119}.

\bibitem{Shaffer2021}
R.~Shaffer, E.~Megidish, J.~Broz, W.-T. Chen, H.~H{\"a}ffner,
  \enquote{Practical verification protocols for analog quantum simulators},
  {\em npj Quantum Information\/}, vol.~7, no.~1, p.~46, Mar 2021,
  \doi{10.1038/s41534-021-00380-8},
  \urlprefix\url{https://doi.org/10.1038/s41534-021-00380-8}.

\bibitem{LHL2019}
E.~Bairey, I.~Arad, N.~H. Lindner, \enquote{Learning a Local Hamiltonian from
  Local Measurements}, {\em Phys Rev Lett\/}, vol. 122, p. 020504, Jan 2019,
  \doi{10.1103/PhysRevLett.122.020504},
  \urlprefix\url{https://link.aps.org/doi/10.1103/PhysRevLett.122.020504}.

\bibitem{HT2020}
Z.~Li, L.~Zou, T.~H. Hsieh, \enquote{Hamiltonian Tomography via Quantum
  Quench}, {\em Phys Rev Lett\/}, vol. 124, p. 160502, Apr 2020,
  \doi{10.1103/PhysRevLett.124.160502},
  \urlprefix\url{https://link.aps.org/doi/10.1103/PhysRevLett.124.160502}.

\bibitem{exact_scalings}
A.~Dembo, A.~Montanari, S.~Sen, \enquote{{Extremal cuts of sparse random
  graphs}}, {\em The Annals of Probability\/}, vol.~45, no.~2, pp. 1190 --
  1217, 2017, \doi{10.1214/15-AOP1084},
  \urlprefix\url{https://doi.org/10.1214/15-AOP1084}.

\bibitem{batagelj2005efficient}
V.~Batagelj, U.~Brandes, \enquote{Efficient generation of large random
  networks}, {\em Physical Review E\/}, vol.~71, no.~3, p. 036113, 2005.

\bibitem{penrose2003random}
M.~Penrose, {\em Random geometric graphs\/}, vol.~5, OUP Oxford, 2003.

\bibitem{silverio2022pulser}
H.~Silv{\'e}rio, S.~Grijalva, C.~Dalyac, L.~Leclerc, P.~J. Karalekas,
  N.~Shammah, M.~Beji, L.-P. Henry, L.~Henriet, \enquote{Pulser: An open-source
  package for the design of pulse sequences in programmable neutral-atom
  arrays}, {\em Quantum\/}, vol.~6, p. 629, 2022.

\bibitem{skopt}
T.~Head, M.~Kumar, H.~Nahrstaedt, G.~Louppe, I.~Shcherbatyi, {\em
  scikit-optimize/scikit-optimize\/}, Zenodo, Oct 2021,
  \doi{10.5281/ZENODO.5565057},
  \urlprefix\url{https://zenodo.org/record/5565057}.

\bibitem{hagberg2008exploring}
A.~Hagberg, P.~Swart, D.~S~Chult, {\em Exploring network structure, dynamics,
  and function using NetworkX\/}, Los Alamos National Lab.(LANL), Los Alamos,
  NM (United States), 2008.

\bibitem{sklearn}
O.~Grisel, A.~Mueller, L.~, A.~Gramfort, G.~Louppe, P.~Prettenhofer,
  M.~Blondel, V.~Niculae, J.~Nothman, T.~J. Fan, A.~Joly, J.~Vanderplas,
  G.~Lemaitre, M.~Kumar, L.~Estève, H.~Qin, N.~Hug, N.~Varoquaux, R.~Layton,
  J.~H. Metzen, J.~Du~Boisberranger, A.~Jalali, R.~(Venkat)~Raghav,
  J.~Schönberger, R.~Yurchak, W.~Li, T.~D. La~Tour, C.~Woolam, K.~Eren, E.~,
  {\em scikit-learn/scikit-learn: scikit-learn 1.1.1\/}, Zenodo, May 2022,
  \doi{10.5281/ZENODO.6563718},
  \urlprefix\url{https://zenodo.org/record/6563718}.

\bibitem{qutip}
J.~Johansson, P.~Nation, F.~Nori, \enquote{QuTiP 2: A Python framework for the
  dynamics of open quantum systems}, {\em Computer Physics Communications\/},
  vol. 184, no.~4, pp. 1234--1240, 2013,
  \doi{https://doi.org/10.1016/j.cpc.2012.11.019},
  \urlprefix\url{https://www.sciencedirect.com/science/article/pii/S0010465512003955}.

\bibitem{manzano2020short}
D.~Manzano, \enquote{A short introduction to the Lindblad master equation},
  {\em Aip Advances\/}, vol.~10, no.~2, p. 025106, 2020.

\bibitem{Henry_2014}
L.-P. Henry, T.~Roscilde, \enquote{Order-by-Disorder and Quantum Coulomb Phase
  in Quantum Square Ice}, {\em Physical Review Letters\/}, vol. 113, no.~2, jul
  2014, \doi{10.1103/physrevlett.113.027204},
  \urlprefix\url{https://doi.org/10.1103%2Fphysrevlett.113.027204}.

\end{thebibliography}

\end{document}